\PassOptionsToPackage{table,usenames,dvipsnames}{xcolor}
\documentclass[11pt]{article}
\renewcommand{\baselinestretch}{1.15} 

\usepackage{fullpage} 
\usepackage{lmodern} 
\usepackage[T1]{fontenc} 
\usepackage{microtype} 
\usepackage{url}
\usepackage{amssymb}
\usepackage{mathtools} 
\usepackage{amsthm} 
\usepackage{bm} 
\usepackage{xcolor} 
\usepackage[square,sort&compress,numbers]{natbib} 
\usepackage[colorlinks=true,pdfborder={0 0
0},citecolor=RoyalBlue,linkcolor=black,urlcolor=Magenta]{hyperref}
\usepackage{cleveref} 
   \crefname{figure}{Figure}{Figures}
   \crefname{table}{Table}{Tables}
   \crefname{theorem}{Theorem}{Theorems}
   \crefname{lemma}{Lemma}{Lemmas}
   \crefname{remark}{Remark}{Remarks}
   \crefname{claim}{Claim}{Claims}
   \crefname{section}{Section}{Sections}
   \crefname{observation}{Observation}{Observations}
   \crefname{note}{Note}{Notes}
\usepackage{multirow}
\usepackage{booktabs} 
\usepackage{array} 
\usepackage{graphicx}
\graphicspath{./figs}
\usepackage{grffile} 
\usepackage{caption}
\usepackage{subcaption} 
\usepackage{enumitem} 
\setlist[itemize]{leftmargin=15pt,labelsep=5pt,noitemsep,topsep=5pt}
\setlist[enumerate]{leftmargin=20pt,labelsep=10pt,noitemsep,topsep=5pt}
\newlist{todo}{itemize}{2}
\setlist[todo]{leftmargin=20pt,labelsep=5pt,noitemsep,topsep=5pt}
\newlist{inline}{enumerate*}{1}
\setlist[inline]{before=\unskip{: }, itemjoin={{; }}, itemjoin*={{; and }}, label={(\roman*)}}
\usepackage[ruled,linesnumbered]{algorithm2e}
\usepackage{etoolbox}
\newtoggle{abstract}

\newcommand{\madhav}[1]{}
\newcommand{\bllcomment}[1]{}
\newcommand{\samarth}[1]{}
\newcommand{\dxcomment}[1]{}

\newcommand{\ttt}[1]{\texttt{#1}}

\DeclareMathOperator*{\argmax}{arg\,max}

{\bfseries}{\itshape} 
{\bfseries}{\itshape}
{\bfseries}{\itshape}
\theoremstyle{definition}

\makeatletter

\newcommand{\Rmnum}[1]{\expandafter\@slowromancap\romannumeral #1@}
\makeatother
\definecolor{Blue}{HTML}{4e79a7}
\definecolor{Orange}{HTML}{4e79a7}
\definecolor{Red}{HTML}{e15759}

\newcommand{\ds}{\mbox{$\mathcal{DS}$}}

\newcommand{\livestock}{\mathcal{L}}
\newcommand{\types}{\Theta_\livestock}
\newcommand{\type}{\theta}
\newcommand{\cattle}{\text{\ttt{cattle}}}
\newcommand{\poultry}{\text{\ttt{poultry}}}
\newcommand{\hogs}{\text{\ttt{hogs}}}
\newcommand{\sheep}{\text{\ttt{sheep}}}
\newcommand{\subtypes}{\Gamma}

\newcommand{\beef}{\text{\ttt{beef}}}
\newcommand{\milk}{\text{\ttt{milk}}}
\newcommand{\other}{\text{\ttt{other}}}
\newcommand{\farms}{\mathcal{F}}

\newcommand{\birds}{\mathcal{B}}
\newcommand{\species}{\Theta_\birds}
\newcommand{\abundance}{A}
\newcommand{\birdnet}{G_\birds}

\newcommand{\hpop}{\mathcal{H}}
\newcommand{\pop}{\pi}
\newcommand{\demoset}{\Delta}
\newcommand{\demo}{\delta}
\newcommand{\employset}{\mathcal{E}}
\newcommand{\employ}{\epsilon}

\newcommand{\proc}{\mathcal{P}}

\newcommand{\wimin}{\mbox{$W_i^{\mathrm{min}}$}}
\newcommand{\wimax}{\mbox{$W_i^{\mathrm{max}}$}}

\newcommand{\cafomaps}{\textsc{CafoMaps}}
\newcommand{\bls}{\textsc{BLS}}

\newcommand{\uspop}{\textsc{USPop}}

\newcommand{\agcensus}{\textsc{AgCensus}}
\newcommand{\glw}{\textsc{GLW}}

\newcommand{\eBird}{\textsc{eBird}}
\newcommand{\reports}{\textsc{H5N1Cases}}

\usepackage{soul}

\usepackage{listings}
\usepackage{xcolor}
\usepackage{framed}

\definecolor{codebackground}{rgb}{0.95,0.95,0.95}
\definecolor{codeborder}{rgb}{0.8,0.8,0.8}
\definecolor{codestring}{rgb}{0.6,0.1,0.1}
\definecolor{codekeyword}{rgb}{0.1,0.1,0.6}
\definecolor{codecomment}{rgb}{0.2,0.5,0.2}

\usepackage{authblk} 

\title{A High-Resolution, US-scale Digital Similar of Interacting Livestock,
Wild Birds, and Human Ecosystems with Applications for Multi-host Epidemic Spread}

\author[1,4]{Abhijin~Adiga\thanks{These authors contributed equally to this work.}}
\author[2]{Ayush~Chopra$^*$}
\author[1]{Mandy~L.~Wilson$^*$}
\author[1]{S.~S.~Ravi}
\author[1]{Dawen~Xie}
\author[1]{Samarth~Swarup}
\author[1]{Bryan~Lewis}
\author[1]{Andrew~Warren}
\author[4]{John Barnes}
\author[2]{Ramesh~Raskar}
\author[1,3,4]{Madhav~V.~Marathe}

\affil[1]{Biocomplexity Institute, University of Virginia, Charlottesville,
VA}
\affil[2]{ MIT Media Lab, Massachusetts Institute of Technology, Cambridge,
MA}
\affil[3]{Dept. of Computer Science, University of Virginia, Charlottesville,
VA}
\affil[4]{Office of Molecular Detection, CDC GA}
\affil[5]{email: \texttt{abhijin@virginia.edu} and \texttt{marathe@virginia.edu}}
\date{}

\begin{document}

\maketitle


\begin{abstract}
One Health issues, such as the spread of highly pathogenic avian
influenza~(HPAI), present significant challenges at the
human-animal-environmental interface. Recent H5N1 outbreaks underscore the
need for comprehensive modeling efforts that capture the complex
interactions between various entities in these interconnected ecosystems.
To support such efforts, we develop a methodology to construct a synthetic
spatiotemporal gridded dataset of livestock production and processing,
human population, and wild birds for the contiguous United States, called a
\emph{digital similar}. This representation is a result of fusing diverse
datasets using statistical and optimization techniques, followed by
extensive verification and validation. The livestock component includes
farm-level representations of four major livestock types -- cattle,
poultry, swine, and sheep -- including further categorization into subtypes
such as dairy cows, beef cows, chickens, turkeys, ducks, etc. Weekly
abundance data for wild bird species identified in the transmission of
avian influenza are included.  Gridded distributions of the human
population, along with demographic and occupational features, capture the
placement of agricultural workers and the general population.  We
demonstrate how the digital similar can be applied to evaluate spillover
risk to dairy cows and poultry from wild bird population, then validate
these results using historical H5N1 incidences. The resulting
subtype-specific spatiotemporal risk maps identify hotspots of high risk
from H5N1 infected wild bird population to dairy cattle and poultry
operations, thus guiding surveillance efforts.
\end{abstract}


\section{Introduction}
\label{sec:background_summary}
Highly Pathogenic Avian Influenza~(HPAI) poses a serious global threat to health,
environment and food security. In the Americas alone, the
unprecedented spread of H5N1 virus clade~2.3.4.4b has led to severe loss of
wildlife~\cite{caliendo2022transatlantic,leguia2023highly,uhart2024massive,puryear2023highly}.
In the US, the incidence among wild birds is widespread.  Large-scale
outbreaks in poultry and dairy cattle threaten food
production~\cite{burrough2024highly,prosser2024using,caserta2024spillover,nguyen2024emergence}.
There have also been several instances of zoonotic
transmissions~\cite{ush5n1humans} through exposure to poultry and cattle, 
which poses a serious pandemic risk~\cite{koopmans2024panzootic}.
This work is motivated by the urgent need for a modeling platform to
understand and respond to HPAI spread accounting for the various agents
that shape and are affected by this phenomenon.

In recent years, several national-scale,
realistic in silico representations of populations, socioeconomic
activities, and built infrastructures have been developed to study complex
phenomena such as epidemiology, emergency response and food security at
fine spatiotemporal
resolutions~\cite{burdett2015simulating,cheng2023maps,van2021challenges,eubank2004modelling,prosser2024using,harrison2023synthetic}.
Here, we refer to such synthetic datasets as \emph{digital similars}. 
They have statistical similarity to real data, but differ from ``digital
twins''~\cite{mihai2022digital,wu2021digital,delgado2019big,pylianidis2021introducing},
which are intended as precise ``living'' replicas of the real-world systems they
represent~\cite{batty2024digital,caldarelli2023role}.
  These realistic data sets are used
for risk assessment and simulation modeling, as evidenced by studies
conducted during the COVID-19 pandemic to analyze the dynamics of
infectious
diseases~\cite{abueg2020modeling,ferretti2020quantifying,hoops2021high,kerr2021covasim,aleta2020modelling,chen2022effective}.

The first large-scale high-resolution digital similars of socio-technical
systems were developed more than two decades
ago~\cite{eubank2004modelling}. Subsequently, many products have been
proposed to address problems in multiple domains, ranging from simple
gridded distributions of populations with demographic attributes, to definitions of
activities and interactions with built
infrastructure~\cite{chen2022effective,lloyd2017high}. Some of these digital similars
focus on modeling the spatial distribution of
livestock~\cite{gilbert2018global,burdett2015simulating,cheng2023maps,bruhn2012synthesized,prosser2024using,humphreys2020waterfowl}.
Of these previous efforts, those on US-scale data sets, including recent ones in the context
of HPAI, focus on a single livestock type. Key data challenges stem from the need to explore and fuse diverse,
often sparse, data sets, which are misaligned in format and spatial
resolution. Methodological challenges in this context include the choice of 
appropriate objectives, assumptions and constraints in the  algorithmic 
formulations in order to achieve realistic representations that are statistically 
consistent with the parent data sets (for example, composition and distribution of 
livestock farms).

\begin{figure}[htpb]
\centering
\includegraphics[width=\textwidth]{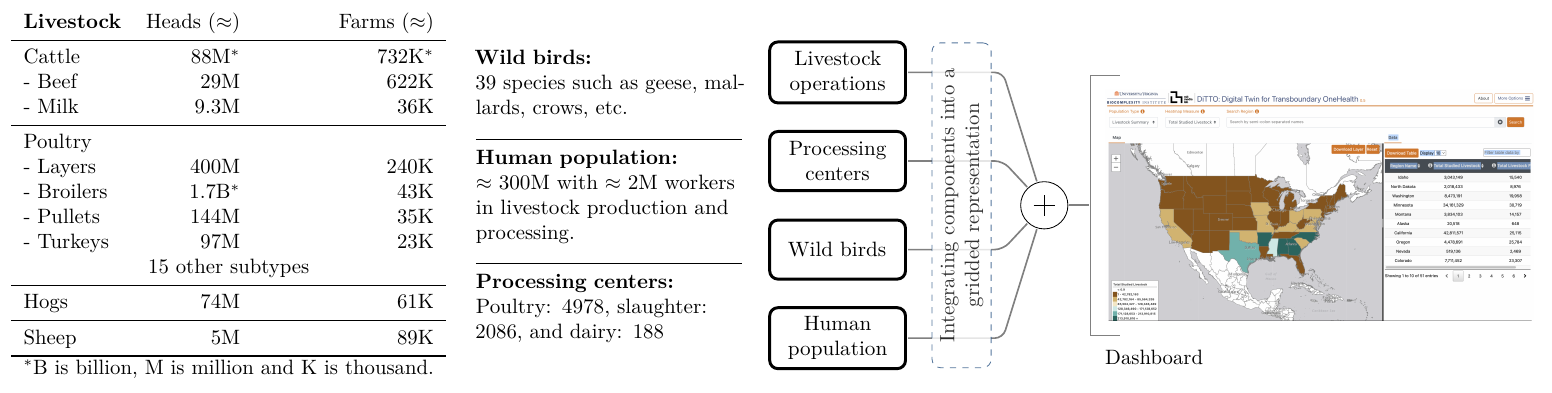}
\includegraphics[width=.95\textwidth]{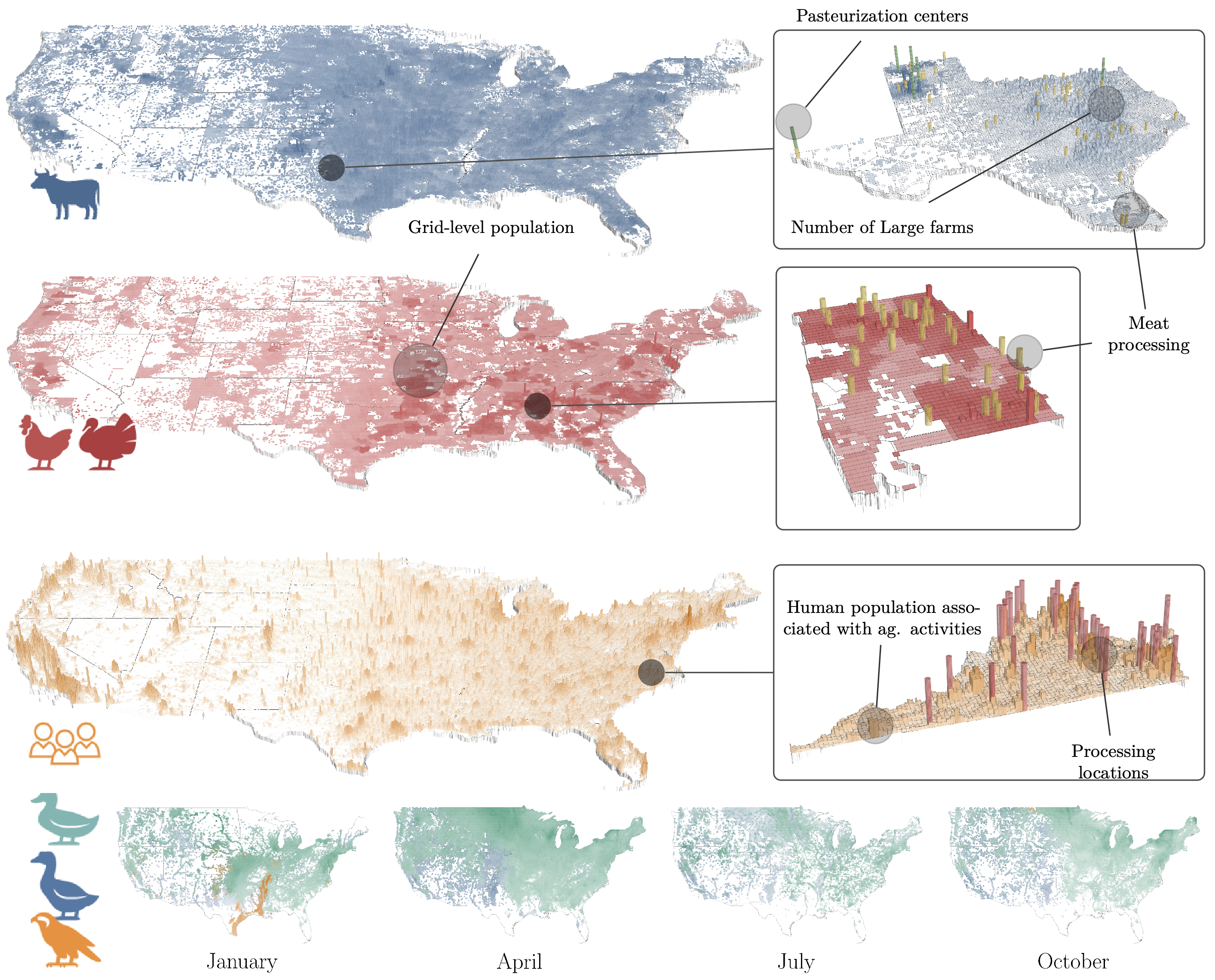}
\caption{\textbf{Overview of the digital similar.} A schematic of the
system highlighting the various components and the dashboard through which
the data is exposed is provided at the top. Two of the four livestock
layers are shown. We have zoomed in on major production regions for the
respective livestock. Both population density and counts of farms are
depicted. Also shown are livestock and dairy processing centers. For the
human population, agricultural workers are highlighted.  The spatiotemporal
distribution of three wild bird populations is shown in the bottom layer.}
\label{fig:ds}
\end{figure}

\paragraph{Summary of our contributions.}
This work presents an approach to develop a high-resolution multi-layered
spatiotemporal representation of the contiguous US, henceforth referred to
as the \emph{digital similar}~(\ds), that captures (i)~the distribution of
livestock populations and operations for multiple types (like cattle or poultry) and 
subtypes (like beef or milk cows, chickens, turkeys, etc.), (ii)~associated food processing
center locations, capacities, and functions, (iii)~spatiotemporally-varying
wild bird abundances for multiple species affected by H5N1, and (iv)~human populations with
demographic features and attributes capturing agricultural employment, as
illustrated in Figure~\ref{fig:ds}. We leverage diverse open datasets
(listed in Table~\ref{tab:data}), such as the Census of Agriculture, the
Gridded Livestock of the World (\glw{}) dataset, \eBird{} Status and
Trends, a digital twin of the US population, and locations of
livestock-related operations obtained from multiple sources.  Data gaps are
addressed by using a combination of statistical tools and mathematical
programming.  Mapping livestock populations to farms and assigning them to
grid cells are cast as optimization problems and solved using integer
linear programs. We perform rigorous data quality checks with reference to
the source data sets, and verification and validation studies using
independent data sets, including known locations of large livestock farms
and H5N1 incidence reports.

To the best of our knowledge, this is the first work to model multiple
livestock types and subtypes on a national scale; previous works have
focused on a single type of livestock~(e.g.,
\cite{prosser2024using,burdett2015simulating,humphreys2020waterfowl}).  We
demonstrate its utility as a comprehensive platform for modeling and risk
assessment of HPAI-like phenomena at high spatial and temporal resolutions,
thus informing disease surveillance and control efforts. We assess the
spillover risk from the H5N1-infected wild bird population to dairy cattle
and poultry. We develop spatiotemporal risk maps for various subtypes,
identifying hotspots for spillover risk and zoonosis.

\begin{table}[htb]
\centering
\rowcolors{1}{white}{gray!15}
\caption{Datasets explored to construct and validate the digital similar.
Throughout the paper, each dataset will be referred to by its abbreviation.
}
{\footnotesize
\begin{tabular}{p{3.5cm}p{1.8cm}p{1.8cm}p{7.2cm}}
\toprule
\textbf{Name} & \textbf{Abbrv.} & \textbf{Source} & \textbf{Description} \\
\midrule
Census of Agriculture & \agcensus & \cite{agcensus2022,agcensus2022download,agcensus_layer} & Provides location-level Ag data such as number of farms, farm sizes, crop types, fallowed status. Provides individual-level Ag data such as number of workers on a farm \\
Gridded Livestock of the World & \glw & \cite{robinson2014mapping,faoGriddedLivestock,gilbert2018global} & GLW4.0 provides distribution maps for several livestock types. FAO hosts this website. \\
eBird Status and Trends & \eBird & \cite{eBirdStatusTrends2022,sullivan2009ebird} & Weekly data of relative abundance of migratory birds across geospatial regions throughout the year. \\
Dairy processing & Dairy plants & \cite{usdadairy} & Large dairy processing centers and their attributes regulated by USDA AMS. \\
Meat, poultry, and egg processing & Meat and poultry & \cite{usdameat} & Listing of establishments that produce meat, poultry, and/or egg products regulated by USDA FSIS. \\
US population & \uspop & \cite{adiga2015generating,chen2025epihiper} & Synthetic digital twin of the US human population \\
Quarterly Census of Employment and Wages & \bls & \cite{bls} & Quarterly counts of livestock workers (NAICS code 112) from Bureau of Labor Statistics \\
CAFOs in the US & \cafomaps & \cite{cafomaps,statecafoguides} & A map of Concentrated Animal Feeding Operations (CAFOs) in the Southern United States covering nine states. \\
H5N1 outbreaks & \reports & \cite{ush5n1birds,ush5n1poultry,ush5n1humans,woahh5n1} & H5N1 bird flu detections in wild birds, livestock, and humans by state and county \\
\bottomrule
\end{tabular}

}
\label{tab:data}
\end{table}

\iftoggle{abstract}{
}{
\newcommand{\bhfive}{\mbox{$B_{\text{H5}}$}}
\section{Results}
\label{sec:results}

\subsection{The Digital Similar}
\label{sec:ds}

The digital similar \ds{} provides a unified gridded
representation of livestock production and processing operations, the human
population, and wild bird populations in the contiguous US.
Figure~\ref{fig:ds} provides a layered view of the digital similar along with
a summary of population sizes. Table~\ref{tab:data} provides an overview 
of the data sources used to construct it. Formally,
$\ds(V,\livestock,\proc,\birds,\hpop)$ is defined over a grid~$V$ overlaid
on the study region, which, in our case, is the contiguous United States.
Each grid cell~$v\in V$ has attributes that capture the details of each of
these components. In the current setting, we use a $5\times5$~arc
minute$^2$ grid.  Descriptions of the components
$\livestock$, $\proc$, $\birds$, and $\hpop$ are provided below.

\paragraph{Livestock
$\livestock\big(\types,\{\subtypes_\type\mid
\type\in\types\},\{\farms_\type\mid \type \in\types\}\big)$.} 
We develop a novel generic approach to construct the livestock layers from
agricultural census and grid-level estimates of livestock populations. 
Figure~\ref{fig:livestock} outlines the methods comprising of statistical 
methods and optimization techniques.
The livestock population comprises four types of
animals: $\types=\{\cattle,\poultry,\hogs,\sheep\}$. For each
type~$\type\in\types$, $\subtypes_\type$ denotes the set of different ``subtypes''
of animals. For example, $\subtypes_\cattle=\{\beef,\milk,\other\}$. The
full list of subtypes is provided in the supplement. For each type of
livestock, the population is partitioned into farms\footnote{All livestock 
production operations will be referred to as farms.}. The
collection of farms for each livestock type~$\type$ is denoted by
$\farms_\type$. For each farm~$f\in\farms_\type$, the population of each
subtype~$\gamma$, denoted by~$H_{f\gamma}$, is specified.  (We use $H$ for
head counts).  Also specified is the grid cell~$v$ to which this farm is
assigned. Note that, in the current digital similar, farms with mixed
livestock types (e.g., farms with both cattle and hogs) are not
represented. Therefore, the sum total of farms across livestock types
would exceed the total number of livestock farms. 

\paragraph{Processing centers $\proc$.} This layer provides
information regarding livestock-associated food processing centers such as meat
processing, dairy processing, and poultry processing units. Each processing
unit~$p\in\proc$ contains attributes such as the location of the unit,
type of processing, and the size estimate. 

\paragraph{Wild birds $\birds\big(\species, \abundance(\cdot)\big)$.} 
This component captures the spatiotemporal distribution of
multiple species of birds identified as significant vectors of avian
influenza based on H5N1 incidence data from 2022--2024. 
There are~36 species of birds represented in this component derived from
\eBird{} data and H5N1 incidence data from 2022--2024~(see~Table~\ref{tab:data}
for the data sources).
Let~$\species$ denote the set of different species (listed in the supplement).
The abundance of a species~$\type\in\species$
in grid cell~$v\in V$ at time~$t$ is denoted by~$\abundance(\theta,v,t)$.
The data is available at a weekly resolution. 

\paragraph{Human population $\hpop\big(\demoset,\employset,\pop(\cdot)\big)$.}
This component provides a grid-level representation of the human population
with emphasis on agricultural workers (workers associated with livestock and its
processing). For each cell~$v$,~$\pop(v,\demo,\employ)$ denotes the size of
the subpopulation that belongs to the demographic group~$\demo\in\demoset$
defined by attributes including age group and sex and employed in
professional classes specified by~$\employ\in\employset$. An employment
group~$\employ$ is defined by occupation and industry attributes, where
non-agricultural employees are all binned into one group, namely
non-agriculture.

\begin{figure}[htb]
\centering
\includegraphics[width=.8\textwidth]{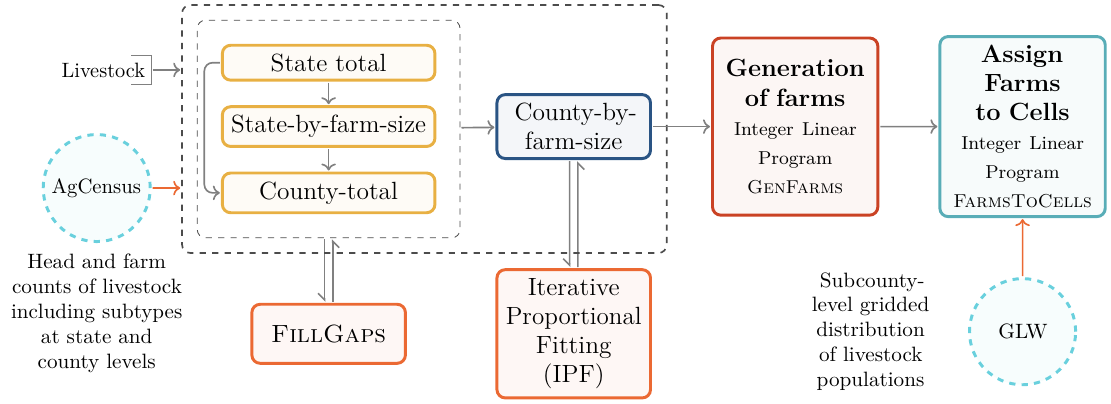}
\caption{A schematic of the livestock layer construction, including
data processing, generation of farms and assignment of cells to farms. It takes 
as input \agcensus{} data that comprises head and farm counts at various
administrative levels and the gridded distribution of the livestock populations
from \glw.  \textsc{FillGaps} is an integer program that fills gaps in the
census data.  \textsc{GenFarms} is an integer linear program~(ILP) for
distributing the livestock populations to farms consistent with the census data.
\textsc{FarmsToCells} is an ILP that assigns farms to grid cells with the
objective of aligning the population with \glw{}.
}
\label{fig:livestock}
\end{figure}

\subsection{Farm generation and cell assignment verification}
Here, we evaluate the construction process for the livestock layers (outlined in Figure~\ref{fig:livestock}) by comparing
the constructed layers with the parent datasets.  We compared the
total head counts from the assigned farms with the corresponding counts
from the \agcensus{} data. The aggregation was done at the state level. The
absolute relative difference is plotted in
Figure~\ref{fig:livestock_head_counts}~(top row).  The differences between
the modeled head counts and AgCensus are caused by the assignment of head
counts to areas where AgCensus head counts were unreported, and by
subsequent adjustment of counts for consistency across farm
sizes~(Section~\ref{sec:fill_gaps}).  We observe that the relative
difference is below~$1\%$ in most instances, barring a few outliers. Also,
the larger the population of a subtype, the smaller the relative
difference.  Figure~\ref{fig:livestock_head_counts}~(bottom row) shows the
distribution of the livestock population into farms. The largest cattle
farms are assigned around 100,000 heads, while chicken farms (corresponding
to subtypes layers and broilers) can be assigned up to 5~million heads.

\medskip 

We now analyze the performance of two constrained 
optimization algorithms (namely,
\textsc{GenFarms} and \textsc{FarmsToCells}) used in this
work.
These algorithms are based on ILP formulations in which
optimization objectives and the constraints are chosen
carefully based on the available data and the necessary
outcomes.
Detailed descriptions of these algorithms are provided
in the methods and supplement.

\begin{figure}[htb]
\centering
\begin{subfigure}[b]{\textwidth}
\includegraphics[width=\textwidth]{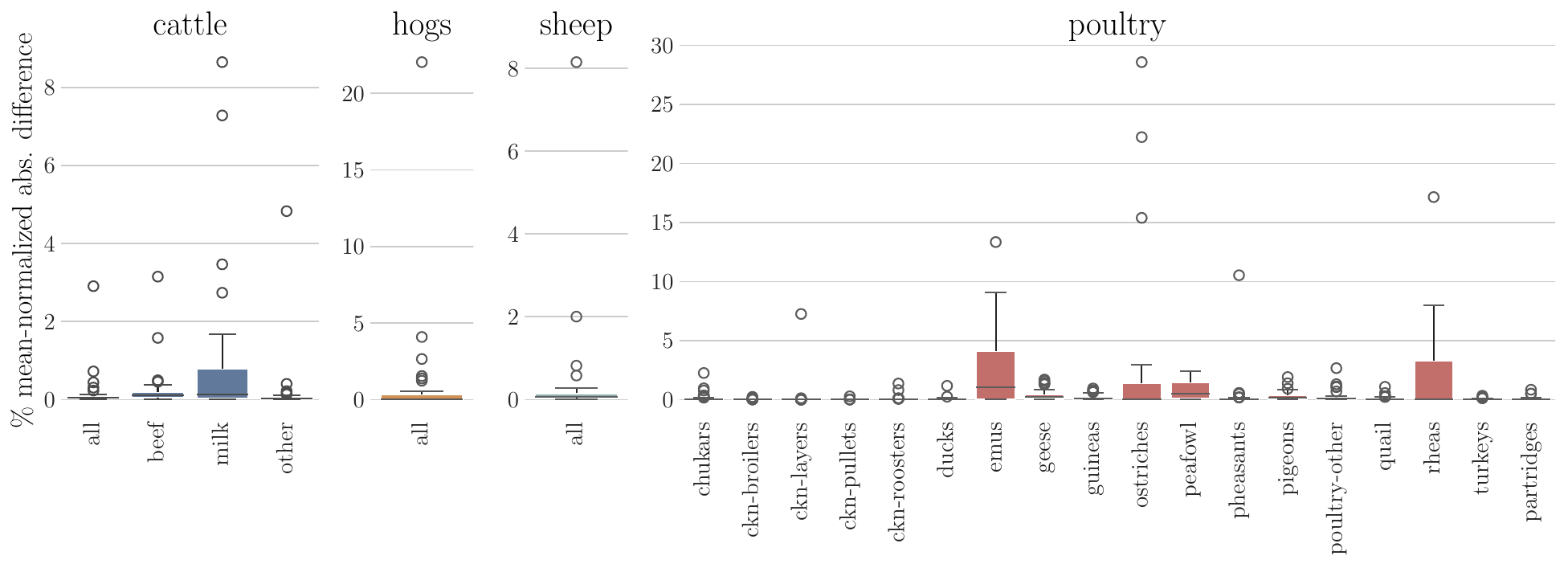}
\caption{}
\label{fig:livestock_head_counts:heads}
\end{subfigure}
\begin{subfigure}[b]{\textwidth}
\includegraphics[width=\textwidth]{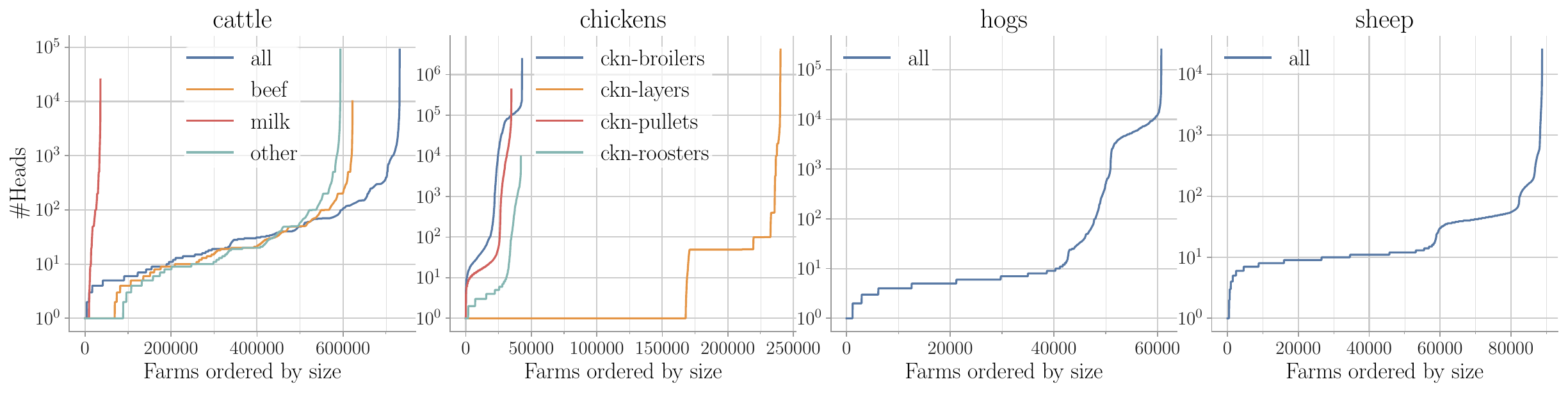}
\caption{}
\label{fig:livestock_head_counts:farms}
\end{subfigure}
\caption{Alignment of the livestock layers with \agcensus{} and
\glw{} datasets. (a)~Head counts of assigned farms are compared with
\agcensus{}. We have a plot for each livestock type with subtypes on the
x-axis and percentage mean-normalized absolute relative difference of state totals from the
census and \ds{} on the y-axis.
(b)~The distribution of livestock populations among farms ordered by farm
size. The y-axis corresponds to farm size. Separate plots for subtypes are
shown for cattle and poultry.
}
\label{fig:livestock_head_counts}
\end{figure}

We first analyze the performance of \textsc{GenFarms}. In this algorithm, 
the minimization objective includes 
the parameter~$\lambda_1$, which
bounds the error in livestock totals by farm
size category between the reported value and our assignment. (This error is
due to the gap-filling step carried out by the IPF process.) Our results in
Figure~\ref{fig:livestock_farms_to_cells}(a) show that this discrepancy is
low across livestock types, which implies that the assignment is close
to the known total head counts. Next, we analyze the performance of
\textsc{FarmsToCells}
in two ways, evaluating how well the assignment of farms aligns with the \glw{}
dataset. In Figure~\ref{fig:livestock_farms_to_cells}(b), we plot the
parameter~$\lambda_5$ (see supplement A4), which  
corresponds to the maximum absolute difference between the head counts
corresponding to our assignment and \glw{} cells. The second plot in 
Figure~\ref{fig:livestock_farms_to_cells}(b)
measures the agreement of our aggregated head counts with GLW data using Pearson's
correlation coefficient.  For all livestock types except poultry, the
correlation is, on average, around~$0.75$. However, there are instances
which are negatively correlated with \glw{}. 
The reason for this behavior is that larger farm sizes make it more difficult to 
align the head counts with cell capacities. In general, poultry distribution is
weakly correlated with \glw.

\begin{figure}[htb]
\centering
\begin{subfigure}[b]{.33\textwidth}
\includegraphics[width=\textwidth]{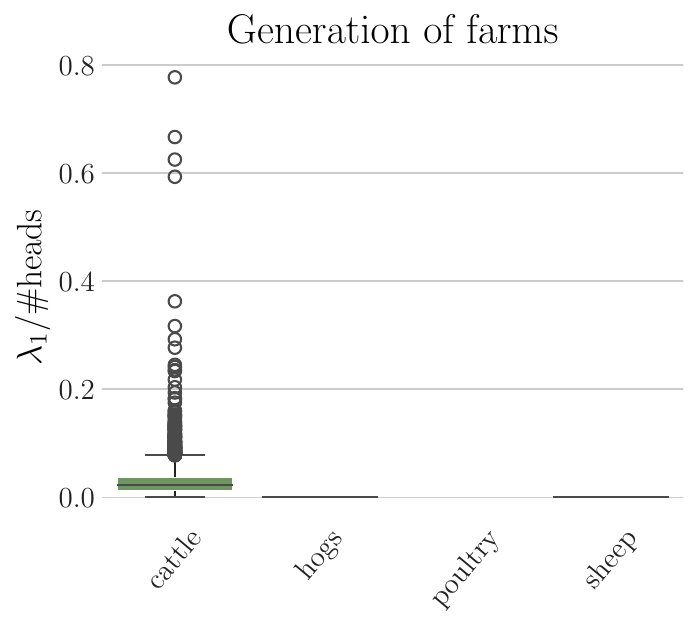}
\caption{\textsc{GenFarms}}
\end{subfigure}
\begin{subfigure}[b]{.62\textwidth}
\includegraphics[width=\textwidth]{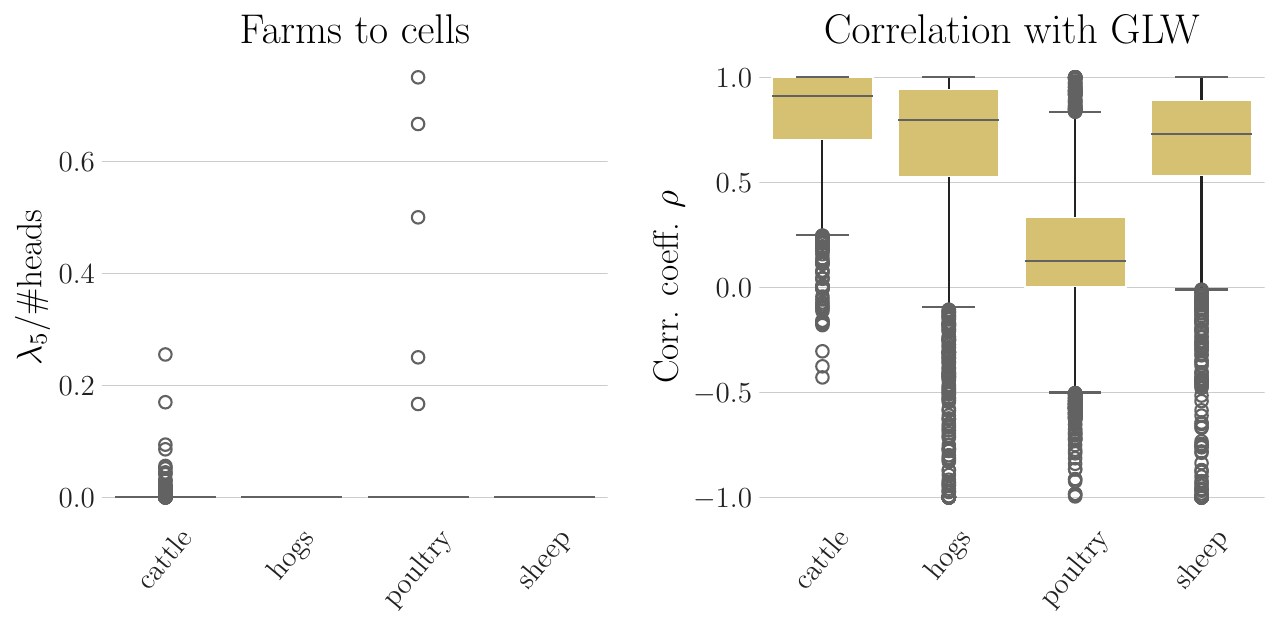}
\caption{\textsc{FarmsToCells}}
\end{subfigure}
\caption{(a)~Analysis of the \textsc{GenFarms} algorithm: We plot the value of
the parameter~$\lambda_1$ (see supplement) relative to the number of heads. This parameter 
is the maximum absolute difference between the number of heads in \agcensus  
~to that in the generated farms at the county level for each farm size
category. Lower is better. Each value in the box plot corresponds to a county. 
We also plot this value for a restricted set of instances where the
county totals are known.  
(b)~Analysis of the \textsc{FarmsToCells} algorithm:
Both plots indicate the agreement of the farm assignment with \glw{} data.
The first plot corresponds to parameter~$\lambda_5$ (see supplement) for county--livestock
instances relative to the number of heads. This parameter is the maximum absolute difference between 
assigned head counts in a cell and its corresponding \glw{} value. Lower is better. Using the Pearson
correlation coefficient, cell-level head counts aggregated from our farm
assignment are compared with \glw{} head counts; higher is better.}
\label{fig:livestock_farms_to_cells}
\end{figure}

\subsection{Validation of farm locations}
Concentrated Agricultural Feed Operations (CAFOs) are large animal feeding
operations that are a potential hazard to the environment and health.
CAFOs are regulated by the Environmental Protection Agency (EPA), and some state agencies provide location information, among other attributes. We obtained such data from
\cafomaps~\cite{cafomaps}. We focus on cattle, hogs, and chickens. For
each county--livestock instance for which such data is present, we
selected large farms from our farm assignment based on 
livestock-specific thresholds informed by CAFO size specifications 
provided by various
states~\cite{statecafoguides}. We computed the Haversine distance of each CAFO location to the
centroid of each grid cell that contains a large farm.  We construct a
weighted complete bipartite graph~$G(A,B)$ for each county--livestock
instance.  Here~$A$ corresponds to farm locations from our assignments;
each farm is assumed to be located at the centroid of the grid cell to which it
belongs.  The set~$B$ corresponds to CAFO locations. For each~$u\in A$
and~$v\in B$, the weight on the edge~$(u,v)$ is the inverse of the distance
between the two locations. 
We compute a maximum weighted perfect matching\footnote{For a complete
bipartite graph $G(A, B, E)$, where $|A| = m$ and
$|B| = n$, a perfect matching consists of
$\min\{m, n\}$ edges.}
of this bipartite graph to match each CAFO location to a
farm in \ds. The main objective is to map as many CAFO locations as
possible. It is possible that the number of farms is greater than the
number of CAFO locations, as not all locations are listed. The results of
the matching are analyzed in Figure~\ref{fig:cafo}. We considered two sets
of thresholds, the second set corresponding to larger farms compared to the
first. A large percentage of CAFO locations were matched in the case of
cattle~($>95$\%) and chickens~($>83$\%), while in the case of hogs we
observe only $50$\% match. A closer examination of \agcensus{} data reveals  the
reason for the low number of matches for hogs: the number of farms specified by the \agcensus{} dataset
for the relevant size categories is less than the number of CAFO locations
specified. Among the matched locations, we observe that $90$\% of the CAFO
locations are at most $10$~miles from the grid centroid of the
corresponding farm from the digital similar, which places it in the same
grid cell or a neighboring one.
\begin{figure}[htb]
\centering
\begin{subfigure}[b]{.58\textwidth}
\includegraphics[width=\textwidth]{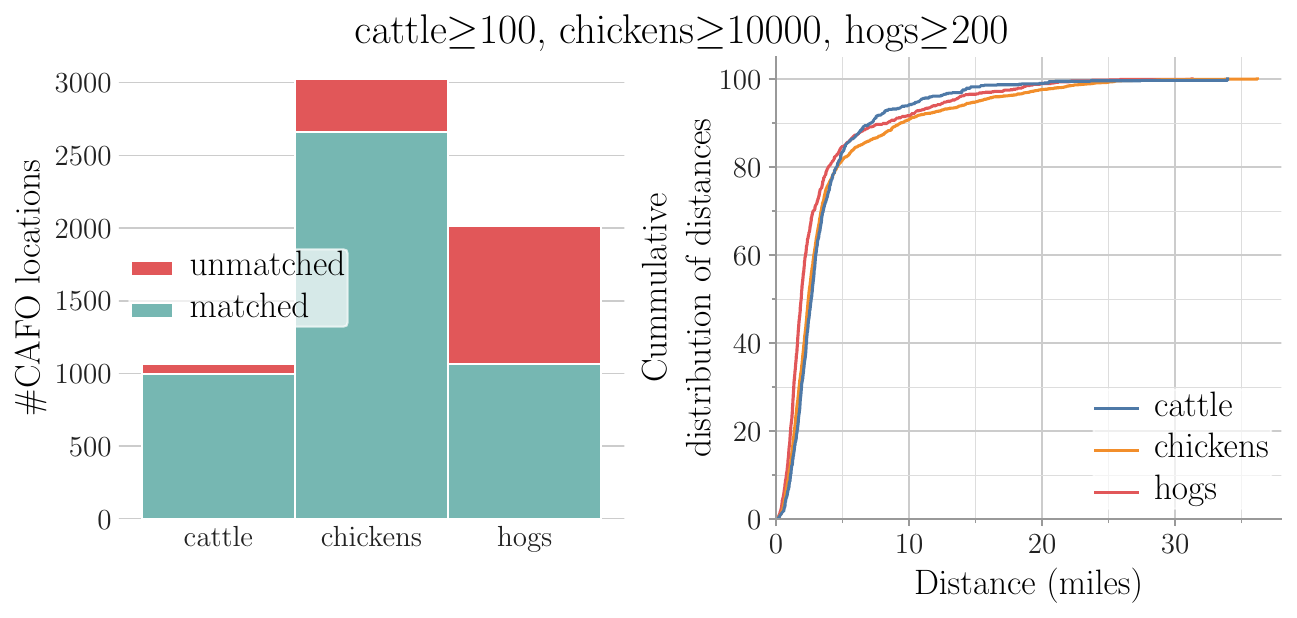}
\caption{}
\label{fig:cafo}
\end{subfigure}
\begin{subfigure}[b]{.38\textwidth}
\includegraphics[width=\textwidth]{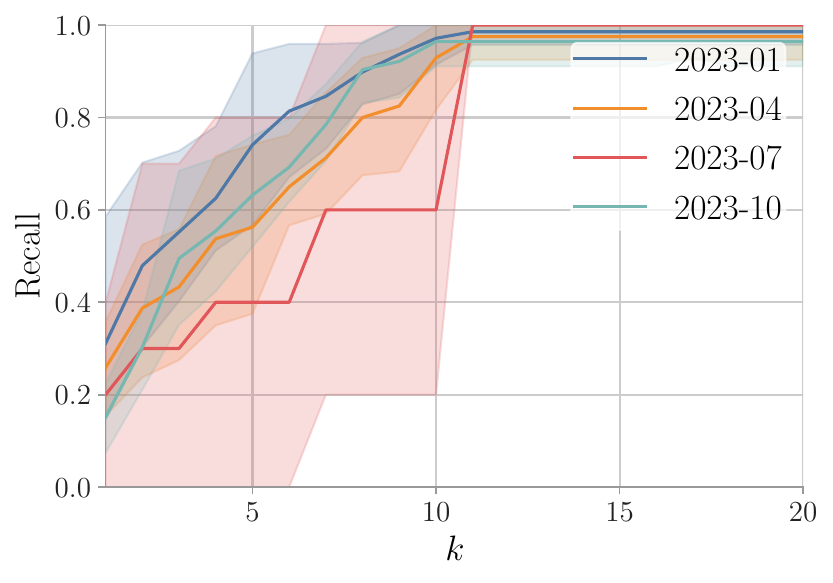}
\caption{}
\label{fig:bird_cases_validation:all}
\end{subfigure}
\caption{Validation of components: (a)~Analysis of mapping CAFO locations
by livestock type to farms from the digital similar. Farms were chosen
based on the thresholds stated in the title. The first subplot shows how
many CAFO locations were matched. The supplement has an additional plot for a different set of thresholds. The second subplot provides the
cumulative distribution of the distances~(in miles) between matched pairs
of CAFO locations and farms. An additional plot for a different set of
thresholds is in the supplement. (b)~Analysis of H5N1 cases and bird
abundance for the period of January to December of~2023. The plot
summarizes the results across eight reporting states for the four quarters. 
The x-axis corresponds to the number of top species groups considered, while the
y-axis corresponds to the count of those groups with H5N1 incidence.}
\label{fig:validation}
\end{figure}
We note that a majority of matched CAFO farms corresponding to cattle and
hogs are within~$20$ miles of the matched farm in \ds.


\subsection{Wild bird abundance and H5N1 incidence}
\label{sec:birds-popln}
We compare the relative abundance of the chosen bird species, as captured
in the \ds, with the occurrence of H5N1 cases at the state level. We recall
that we chose to include all species with reference to H5N1 incidence
data from 2022-2024. The objective is to ascertain whether
there are H5N1 incidences among the most abundant birds in each state where
cases have been reported. This exercise establishes the relevance of the included
bird population to the collocation-based risk
analysis.  For each state, we calculate the average abundance for each
bird species over the study period~(January 1 to March 30, 2023) breaking
down the period into four quarters. We then compile H5N1 case counts for
each bird species in the same state and time frame. Bird species are
grouped into~15 categories (e.g., `Duck', `Goose', `Eagle') based on species similarity to provide more robust comparisons. We employ the \textbf{top-K recall} metric to quantify the relationship 
between bird abundance and H5N1 cases. It is the proportion of species
with H5N1 cases that are among the~$K$ most abundant species in a state 
within the target period and administrative region.
\[
\text{Top-K Recall} ~=~ \frac{\text{
 Species with H5N1 cases in top-K abundant species}} {\text{Total species
 with H5N1 cases}}.
\]
The results presented in Figure~\ref{fig:bird_cases_validation:all}
summarizes the Top-K recall results for all states and different parts of
the year. We observe that, for all quarters, the larger the bird
population, the greater the likelihood of observed H5N1 cases.  Secondly,
we observe that, in most cases, the top~10 abundant birds cover all
reported H5N1 cases~(for~$k\ge10$), so the recall value is close to~$1$.
More plots are shown in the supplement for multiple US states with details
about specific bird types.  These results suggest a robust correlation
between bird abundance and H5N1 case occurrences.

\subsection{Livestock worker population}
\label{sec:pop}
Here, we compare the livestock worker population in the \ds{} with counts
obtained from the Quarterly Census of Employment and Wages data corresponding 
to the year 2023, made available by the Bureau of Labor Statistics (BLS)
(Table~\ref{tab:data}, \bls). To this end, we choose the population associated
with livestock-related occupations~(SOCP~4520XX) or industry~(NAICS~112).  Also,
considering that this population is not time varying, we analyze \bls{} for
seasonal variations in the livestock worker counts. We note that BLS only
counts workers covered under unemployment insurance, due to which farm
owners, self-employed workers, and many workers (e.g., undocumented workers) 
are excluded from the count.

We recall that the livestock worker distribution in the \ds{} is derived
from a digital twin of the US population (Table~\ref{tab:data}, 
 \uspop). This
distribution is imputed from the American Community Survey~(ACS) 5-year
Public Use Microdata Sample~(PUMS). The total count of livestock
workers in the \ds{} is 704,126, while the total number of such individuals in
the PUMS data is 42,233, which is roughly~6\% of the total worker
population. This is consistent with the fact that the PUMS represents
approximately five percent of the US population.

The plot in Figure~\ref{fig:pop_vnv:diff} shows the difference between our data
and the \bls ~data in counts of
livestock worker population by county and state, respectively, for~2769
counties that are common to both the data sets. We note that,
for more than 96\% of the counties, the synthetic population counts exceed
that of \bls. This is expected as \bls{} excludes a significant population
of farm workers as mentioned above. In many cases, the count is zero. However, there are around~105 counties
for which the \ds{} population is less than that of BLS. However, the
difference in this case is usually very small compared to many of the
remaining instances where the counts in \ds{} far exceed \bls. It is
possible that a significant portion of the farm worker population in these counties did not participate in the census.

Figure~\ref{fig:pop_vnv:workers_farms} shows a comparison between
county-level farm counts and the livestock worker population. Due to missing
information about mixing livestock in farms, our total farm count is higher than the total
mentioned in \agcensus. Hence, we only considered farms with head counts of at 
least~100. Generally, for counties with
higher farm counts, the number of workers is higher. But there is a wide
spread in the number of workers for a fixed farm count. Since counts can
depend on farm sizes and livestock types, without additional information
it becomes almost impossible to compare the two quantities in further detail. 

Analysis of \bls{} shows little variation in the county-level counts of
livestock workers across the year. In Figure~\ref{fig:pop_vnv:seasonality},
we have plotted a scatter plot of the coefficient of variation for the four
quarters of year~2023 with respect to the mean number of workers in the
county. With the exception of two outliers, counties with very large
coefficients of variation have very few livestock workers. For the two
outlier counties, there is at least one quarter with zero count, which
could be attributed to missing data.

\begin{figure}[htpb]
\centering
\begin{subfigure}[b]{.32\textwidth}
\includegraphics[width=\textwidth]{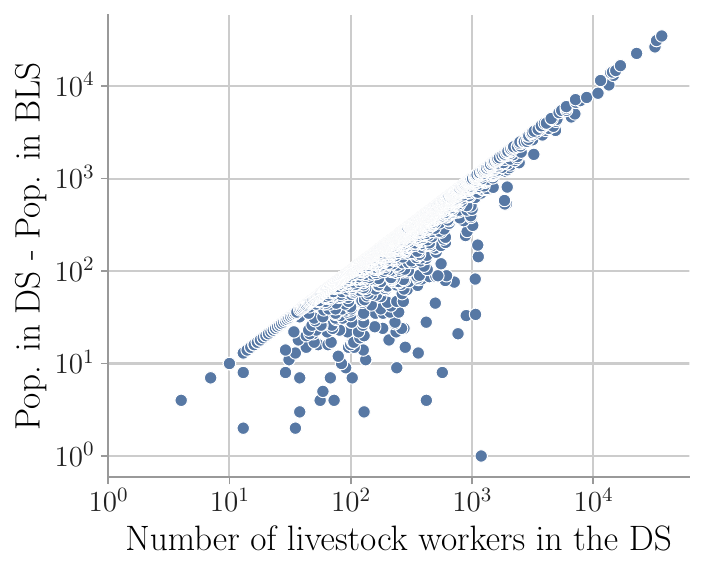}
\caption{Comparing counts with \bls.}
\label{fig:pop_vnv:diff}
\end{subfigure}
\begin{subfigure}[b]{.32\textwidth}
\includegraphics[width=\textwidth]{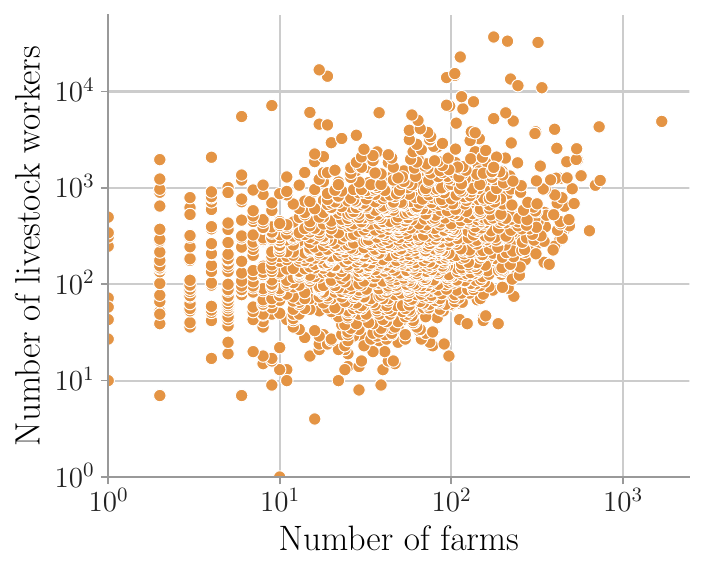}
\caption{Workers vs. farms}
\label{fig:pop_vnv:workers_farms}
\end{subfigure}
\begin{subfigure}[b]{.32\textwidth}
\includegraphics[width=\textwidth]{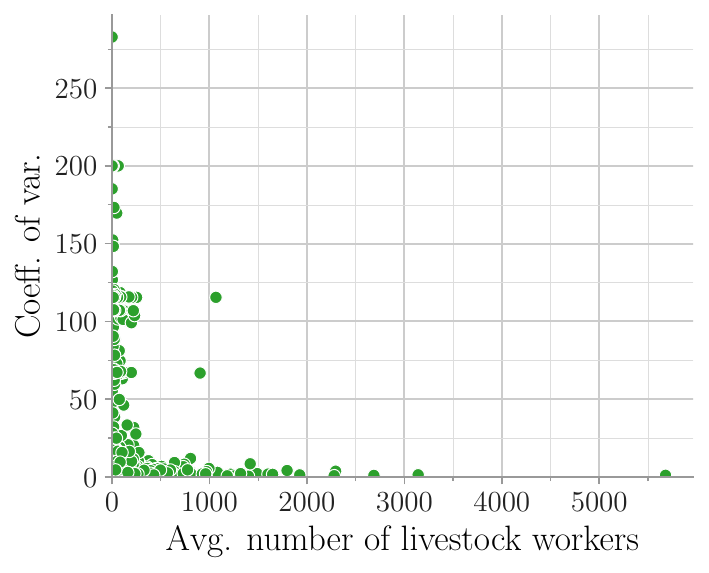}
\caption{Seasonal variations based on \bls.}
\label{fig:pop_vnv:seasonality}
\end{subfigure}
\caption{Analysis of the livestock worker population. All counts presented are
at the county level. (a)~Comparing livestock worker counts between the
digital similar and data from the Bureau of Labor Statistics~(\bls{}).
(b)~Number of livestock workers vs. number of farms at the county level.
(c)~The variation in worker populations across the year. The coefficient of
variation on the x-axis is computed across the four quarters of year 2023.}
\label{fig:pop_vnv}
\end{figure}

\subsection{Risk Estimation}
To demonstrate the utility of this digital similar, we use it to study the
risk of H5N1 spillover from wild birds to various livestock
populations. To this end, we use the livestock and wild bird abundance
layers to conduct simple collocation-based risk assessments at the county
and state levels. Given a livestock subtype~$s$, a grid cell~$i$ and
time~$t$, the risk of spillover to the population of~$s$ is given by
\begin{equation}
    R(i, s, t) = P(i, s) \cdot A(i, t) \cdot \bhfive(i, t),
\end{equation}
where~$P(i,s)$ is the livestock population from the livestock
layer~$\livestock$, $A(i, t)$ is the wild bird abundance at time $t$ from the
corresponding layer~$\birds$ and $\bhfive(i, t)$ is an estimate of the
proportion of the wild bird population infected with the
disease at time $t$~\cite{kent2022spatiotemporal,prosser2024using}.  For each time
period, we aggregate the risk across all grid cells of a county to obtain
risk~$R_c(s,t)$ for a county~$c$. We assign risk percentile ranks to counties, 
designating them as ``Very high''~($\ge 95$ percentile), ``High''~(90--94), 
``Medium''~(75--89) and ``Low''~(0--74). We perform a quarterly assignment. 
Our results are described below. From a livestock perspective, we focus on milk cattle,
turkeys and chicken layers, which have been among the most affected among
the livestock subtypes.

\paragraph{Strong predictive accuracy validates risk assessment framework.}
For both milk cattle and poultry, the risk estimates show strong concordance with
observed H5N1 outbreaks. Figure~\ref{fig:recall} shows results for milk
cattle and turkeys respectively for each quarter. In
general, across subtypes, a large portion of the H5N1 incidences occur in
the very high and high risk counties corresponding to the period of
occurrence demonstrating wild birds as the primary driver of introduction
events in livestock. While our collocation model explains a majority of the incidences, there
is scope to improve this model by giving careful consideration to the functional 
form of the model and accounting for other pathways of spread. The objective here
was to demonstrate the importance of the subtype-specific farm abundance and the abundance of wild birds as some of the main driving factors of this phenomenon.

\begin{figure}
    \centering
    \begin{subfigure}[b]{\textwidth}
        \centering
        \includegraphics[width=.48\textwidth]{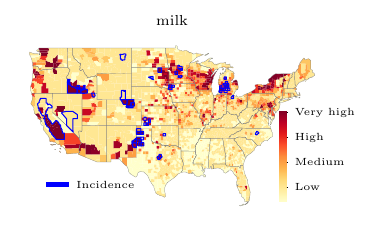}
        \includegraphics[width=.48\textwidth]{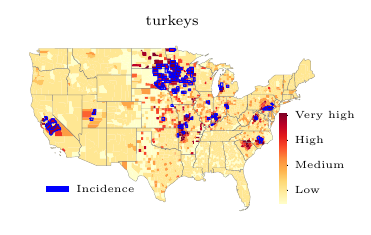}
        \caption{Persistence of elevated risk}
        \label{fig:persistence}
    \end{subfigure}
    \begin{subfigure}[b]{\textwidth}
        \centering
        \includegraphics[width=.48\textwidth]{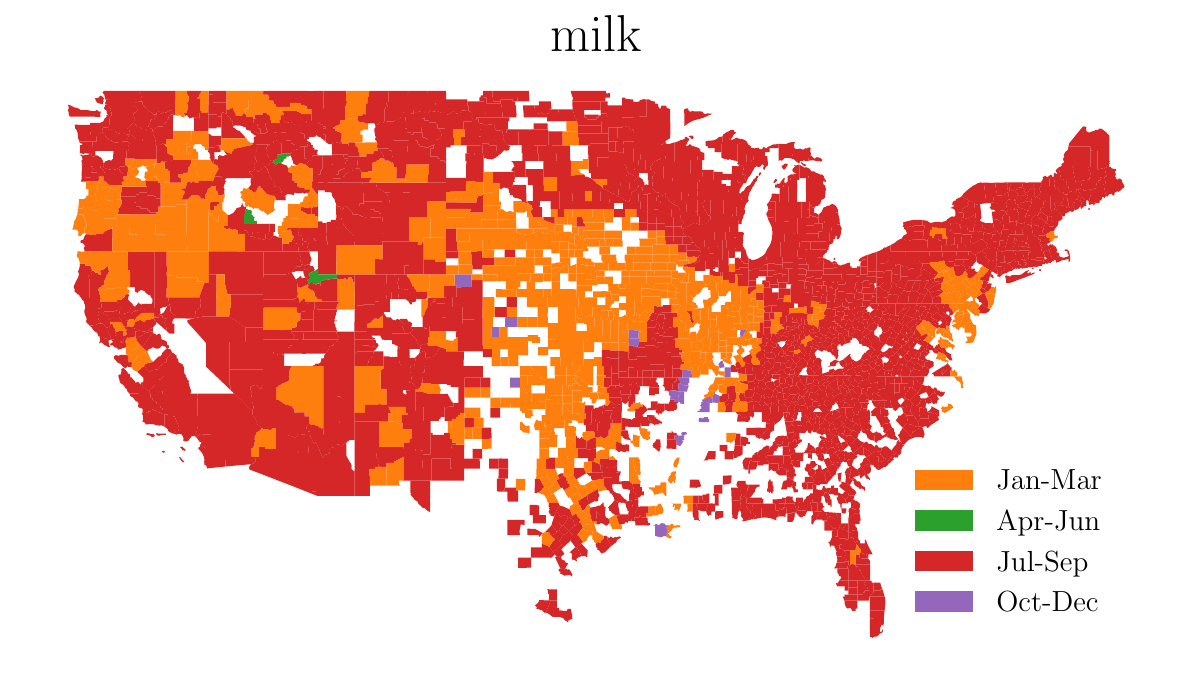}
        \includegraphics[width=.48\textwidth]{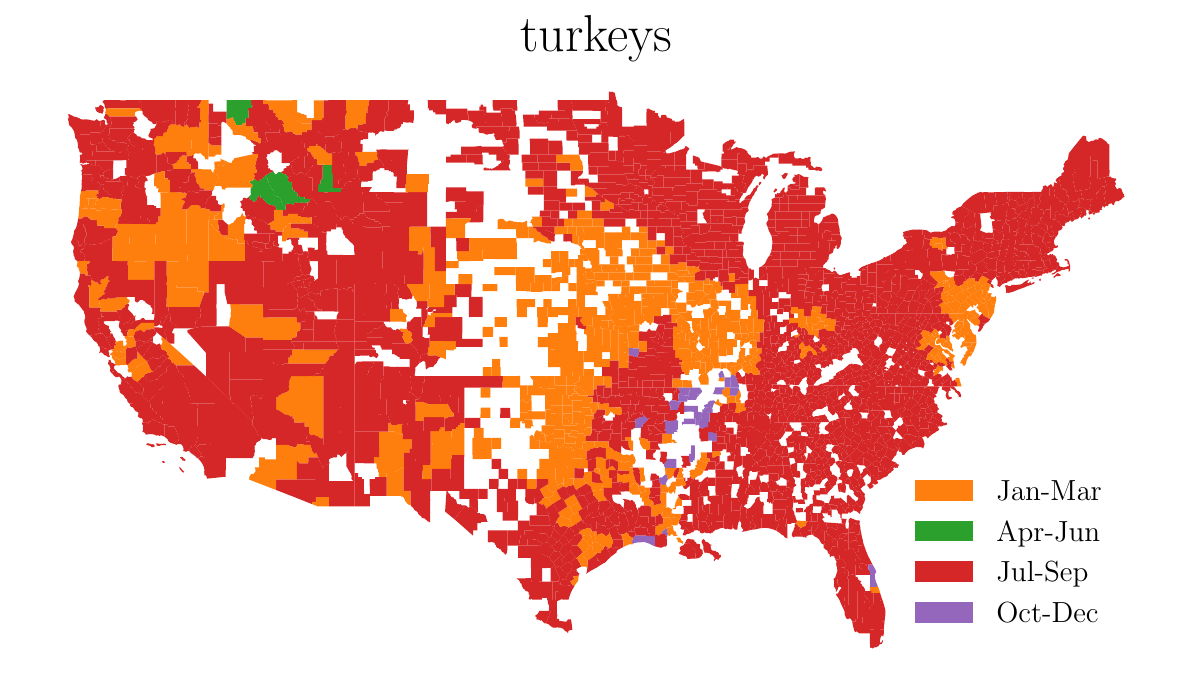}
        \caption{Period of peak risk}
        \label{fig:peak}
    \end{subfigure}
    \begin{subfigure}[b]{.48\textwidth}
        \includegraphics[width=.48\textwidth]{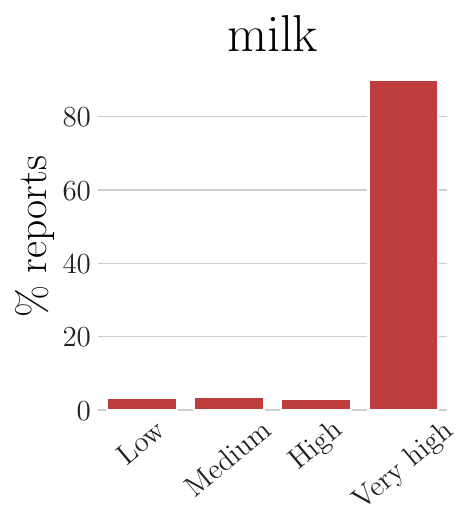}
        \includegraphics[width=.48\textwidth]{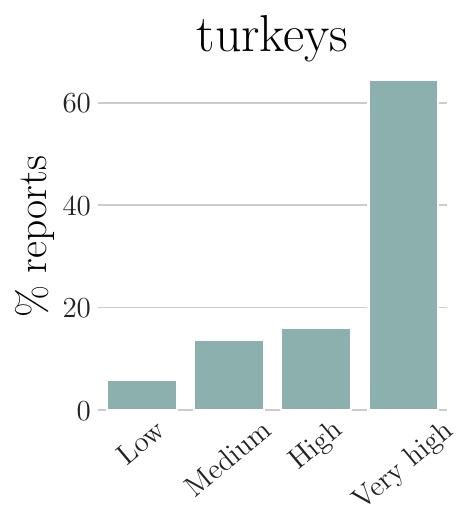}
        \caption{Recall performance}
        \label{fig:recall}
    \end{subfigure}
    \begin{subfigure}[b]{.48\textwidth}
        \centering
        \includegraphics[width=\textwidth]{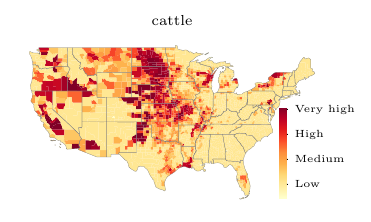}
        \caption{Potential risk to all cattle}
        \label{fig:cattle}
    \end{subfigure}
\caption{Results corresponding to livestock subtype-specific quarterly risk
maps based on county ranks. The risk scores were ranked per quarter and
counties were assigned risk profiles: ``Very high'' ($\ge95\%$), ``High''
($90-94$), ``Medium'' ($75-89$) and ``Low''. 
Here, we show summary plots for milk
cattle and turkeys. (a)~Persistence of elevated risk: Quarter-specific ranks
were combined to obtain a single ranking of counties by sorting counties by
number of occurrences of ``Very high'' risk profiles across quarters
followed by number of occurrences of ``high'' risk profiles and so on.
(b)~Period of peak risk: For each county, we plot the period for which the risk 
is maximum ($\argmax_{t={1,2,3,4}} R_c(s,t)$). (c)~Comparing ground-truth 
incidence reports with risk profiles. Each
incidence is mapped to the risk profile of the corresponding
county--quarter pair. Then, the incidences are binned into the 
respective risk profiles. (d)~Persistence of elevated risk under a potential scenario 
of H5N1 spread in cattle (definition same as in (a)).}
\label{fig:risk_results}
\end{figure}

\paragraph{Risk is subtype specific.}
Figures~\ref{fig:persistence} and~\ref{fig:peak} show the
differences in the spatio-temporal risk across livestock subtypes.
Quarterly risk maps in the supplement better illustrate this difference.
While wild bird abundance generally influences spillover risk, not all
subtypes are affected in the same way as observed in the incidence reports.
This could be driven by biology, farming practices~(e.g., indoor or outdoor
facility), etc. The subtype population also drives the risk factor.
However, we observe that some livestock-intensive counties show up as very
high risk across subtypes informing cross-species spillover risk, as well
as increased exposure to the livestock worker population, which is
typically large in such instances, as per our analysis in
Figure~\ref{fig:pop_vnv:workers_farms}.


\paragraph{Persistent elevated risk informs surveillance priorities.}
Figure~\ref{fig:persistence} maps risk persistence in milk cattle across
counties, highlighting areas that are at elevated risk for multiple time
periods. For both milk cattle and turkeys, there are counties that
consistently rank among the top counties in the very high risk category
over all four periods~(including multiple counties in California, and Weld
County in Colorado) that require constant surveillance throughout the year.
We also observe counties whose rank fluctuates widely~(such as
Lancaster~County in Pennsylvania), requiring time-dependent
surveillance~(results tabulated in the supplement). While California has
the highest risk, for some states like Colorado, this rank changes over time.

\paragraph{Periods of high risk.}
County-level temporal analysis of risk reveals spatial clustering.  The
plots in Figure~\ref{fig:peak} show, for each county, the quarter of
highest risk. For most regions, either the first or the third quarter 
corresponds to the highest levels of risk. Generally, the risk is
spatially clustered, with several county-clusters having the same temporal
risk profiles; however, variations in habitats and the wild bird species hosted can lead 
to differences within a region. Spatial clustering of very high risk
counties increases the potential for multiple spillover events, which
amplify the probability and size of local outbreaks due to other pathways
of spread. 

\paragraph{Future scenarios reveal potential shifts in the risk landscape.}
To evaluate potential future scenarios, we extended our analysis to
consider the spillover risk to the entire cattle population not currently
experiencing H5N1 outbreaks (e.g., among beef cattle).
Figure~\ref{fig:cattle} illustrates how risk hotspots would shift
substantially under this scenario, with increased risk in regions like
North Dakota, Texas, and Kansas, compared to the milk-centric risk maps.
This analysis highlights the importance of early containment and the
potential for broader agricultural impacts.




\section{Discussion}
\label{sec:discussion}


\paragraph{Outline.} Our work presents a comprehensive methodology to
construct synthetic  spatiotemporal datasets of food systems, human
population, and wildlife. To this end, we bring together 
diverse data sets and combine them using combinatorial optimization 
and statistical methods to develop a multi-layered digital similar. 
This is complemented by extensive verification and validation studies.
While the choice of agents and design decisions were influenced by the
focus application -- the
spread of HPAI-like diseases -- the utility of such a digital similar
extends beyond this domain. The following discussion elaborates on 
these points while also highlighting the limitations of the digital
similar.

\paragraph{Related work.}
Several models have been proposed for subcounty-level disaggregation 
of livestock populations. Using a random
forest model, Gilbert~et~al.~\cite{gilbert2018global} develop a global
distribution of populations of multiple livestock types. 
Burdett~et~al.~\cite{burdett2015simulating} develop the FLAPS model to
simulate populations and locations of individual farms for swine
using the \agcensus~dataset and a microsimulation model; this work has been
used to analyze the spread of porcine deltacoronavirus in the
US~\cite{paim2019epidemiology}. 
Cheng~et~al.~\cite{cheng2023maps} develop
the MAPS model to map swine production in China using enterprise
registration information and other datasets. For mapping
concentrated animal feeding operations~(CAFOs), several deep learning
methods have been proposed to map industrial operations using remote
sensing data~\cite{handan2019deep,robinson2022mapping}. In the context of
HPAIs, there is a need for significant extensions to these works given the
requirement to simultaneously account for multiple livestock types and wild
birds, as well as organizational-level distributions. In a very recent
work, Prosser~et~al.~\cite{prosser2024using} address estimation of
transmission risk at the wild waterfowl--domestic poultry interface. They
develop a spatiotemporal model combining~10 species-level wild bird
abundance models~\cite{sullivan2009ebird} with a commodity-level poultry
farm model~\cite{patyk2020modelling}.
They perform phylogenetic analyses to identify
wild bird spillover events to validate their model at two spatial
scales, namely grid-level and county-level.
Humphreys~et~al.~\cite{humphreys2020waterfowl} use a variety of datasets,
including \glw{}, to model waterfowl movement and interactions with poultry
farms and human populations. Our work utilizes several ideas, methods 
and data sets from some of these works to build the digital similar and
for risk assessment.

For many regions, as demonstrated in our risk analysis, colocation of
large livestock populations with wild birds is a good explanation for 
spillover risk. Previous works in the case of poultry~\cite{prosser2024using,humphreys2020waterfowl}
highlight this aspect for the case of poultry. Our results for dairy are
similar in that aspect. However, there are some differences between 
outbreaks in poultry and dairy cattle. While poultry outbreaks have been
largely sporadic, infections in dairy cattle has persisted and spread to
neighboring counties in many states (like Colorado and Michigan in the 
beginning, and later, California). Capturing this would require
knowledge of farm-to-farm movement of animals~(as shown by 
Nguyen~et~al.~\cite{nguyen2024emergence}) and role of agricultural
workers from the perspective of human-mediated pathways, and information 
on spillover from livestock to wild birds and their movement for 
natural pathways. The approaches used in  
BirdFlow~\cite{fuentes2023birdflow} model the flight paths of wild birds, 
but incorporating this into our work would require sample trajectory data.

\paragraph{Limitations.}
While this dataset offers valuable insights as demonstrated in our work, it
is important to acknowledge its limitations and the potential for future
improvements. The dataset faces challenges primarily stemming from the
nature of its parent datasets. For example, the synthetic dataset \glw{} is
misaligned in time with respect to \agcensus. In addition, as shown in our
analysis with respect to farms whose locations are known, not all
assigned operations are matched, indicating spatiotemporal misalignment with
\agcensus{}. Unlike some previous
works~\cite{burdett2015simulating,prosser2024using}, we do not produce
coordinate-level assignments for operations. While this might become a
limitation for very fine-grained analyses, such as farm-to-farm movement of
animals, for such datasets to be useful, additional location- and
operation-level information and rigorous validation is required.  Another
important limitation of our work is that it does not model mixtures of
different livestock types (such as cattle and poultry in the same farm).
For this reason, the total farm count is higher in the \ds{} compared to
\agcensus, particularly in the case of poultry.
For some livestock types (like poultry), there is not enough information about farm sizes, 
leading to a heavy-tailed distribution of populations across farms. While the current openly
available \agcensus{} data does not provide this information, further exploration
of cross-tabulation data that is available by request could allow us to improve on
these aspects. In those instances where even state and county livestock totals are
absent, we fill those gaps based on an equitable distribution of the
missing populations (details in Methods). These adjustments might not be represented in the ground truth.
For wild birds, \eBird{} status and trends data provides only relative
abundance measures, which are subject to observational biases and tend to
underestimate true populations. 
Mapping agricultural workers to farms is a challenging task as it
is a function of farm size, livestock type, and the level of automation
employed~\cite{macdonald2009transformation}. The comparison exercise with
data from BLS highlights the challenges of estimating farm worker counts. 
Despite these limitations, our dataset
provides a valuable foundation for studying complex interactions between
livestock, wild birds, and human populations in the context of avian
influenza transmission.

\paragraph{Conclusion.} Beyond the study of H5N1, the dataset offers valuable
applications to other One Health issues and beyond. The modular nature of
the digital similar enables us to leverage subsets of layers depending on
the nature of the application. This dataset can be applied to model the
spread of other pathogens, such as West Nile virus or Salmonella, which also
involve interactions between livestock and human
populations~\cite{xiao2006semi, westnile2019spatiotemporal,
livestockzoonoses}. Such systems have value beyond infectious diseases in
domains such as food safety, agricultural economics, environmental damage,
pollution, disaster response, biosecurity, and supply chain
problems~\cite{burdett2015simulating,cafomaps,zhu2024high,gilbert2018global}.
In biodiversity and conservation efforts, the wild bird abundance data can
aid in identifying critical habitats and migration corridors, particularly
in the context of livestock operations. Spatially explicit synthetic
datasets are being extensively developed for such non-epidemiological
settings~\cite{meyur2022ensembles,thorve2023high,yuan2023synthetic,barrett2013planning,marathe2014prescriptive}. Our digital similar can extend such works to account for additional
ecosystems such as livestock in the respective applications.

\section{Methods}
\label{sec:methods}

\subsection{Datasets}
Table~\ref{tab:data} summarizes all data used. We used publicly available 
datasets, which can be categorized into three types: census,
synthetic realistic datasets derived from models and data samples, and real
location-level datasets. From a spatial unit perspective, some data (e.g.,
\agcensus{} and H5N1 incidence data) are specified at various administration
levels (county or state), some data  (e.g., \glw{} and \eBird{}) are specified at
the grid level, while exact locations are provided for the rest. Some of these
datasets have been used for the construction of the digital similar, while the
rest have been used for subsequent analysis. More details about each dataset are
provided in the relevant sections.

\subsection{Livestock}
\label{sec:livestock}

Here, we provide an
overview of the process for generating the livestock layers. Two data
sources were used to construct the livestock layers: Census of Agriculture
(\agcensus{}) and Gridded Livestock of the World (\glw{}); these are
described in more detail below.  An overview of the methodology is shown in
Figure~\ref{fig:livestock}.  The types of livestock covered in this work
are cattle, poultry, sheep and hogs; the definitions of livestock type and
subtype are described in the model description in Section~\ref{sec:ds}.
Additional details not covered in this section are provided in the
supplement.

\paragraph{Data organization and availability challenges.}
\agcensus{} provides counts of heads (i.e., population size) and 
farms for various livestock types and subtypes. The data is available at
three different administrative levels -- country, state, and county. The data organization is livestock-type specific,
making it a non-trivial task to extract relevant
information. Farms are
binned into categories based on the head counts of the corresponding
livestock type. We preprocessed the data to ensure that these categories
are disjoint and the categorization is identical at both state and county
levels. Ideally, given a livestock subtype and administrative level, both
farm and head counts are provided for each farm size category. Also
provided is the total count of farms and heads. Together, we have four
possible types of counts: (i)~state-total: total number of farms/heads; 
(ii)~state-by-farm-size: number of farms/heads per farm size category, and
corresponding county-level counts;  (iii)~county-total; and
(iv)~county-by-farm-size. More details with examples are provided in the
supplement. However, some head counts are missing in all count types.
In the case of poultry, even farm size categories are missing for all
subtypes except for (chicken) layers.
The \glw~\cite{faoGriddedLivestock} dataset
provides a gridded distribution of livestock abundance at~$5$~arc minute
resolution for the livestock types, but not for the subtypes. The details of the processing of the \glw{} data are provided in the
supplement.

\paragraph{Filling gaps in data.} We use a combination of integer linear
programs~(ILPs) and iterative proportional fitting~(IPF), the latter following
Burdett~et~al.~\cite{burdett2015simulating}. For cattle and poultry, gaps
are filled for each subtype, while for hogs and sheep, they are filled for
the livestock type.  We use the integer program Algorithm~\ref{alg:fillgap}
(described in the supplement) to fill in  missing data for the following types
of counts: state-total, state-by-farm-size, and county-total. It takes as
input all the known counts, the sum of all the counts, and the bounds on the
unknown counts, and distributes the heads that are unaccounted for equitably 
across all entities for which the counts are missing. The algorithm respects the bounds
provided as input. To fill gaps for county-by-farm-size counts, we follow
the methodology of Burdett~et~al.~\cite{burdett2015simulating}. They apply
IPF~\cite{fienberg1970iterative,deming1940least} to
estimate counts for hogs. At each step, the objective is to make use of all
available data (in all count types). Since subtypes are processed 
independently, the resulting counts can lead to infeasible instances. The
farm generation ILP, \textsc{GenFarms}~(described later), handles such cases.

\paragraph{Farms to cells.} Here, given a livestock type and county, the
objective is to obtain a grid-level distribution of farms that is
consistent with the \agcensus{} data from the perspective of operations and
their sizes, and the \glw{} data from the perspective of the grid-level
distribution of livestock populations. We use a two-step procedure using
optimization algorithms: (i)~generating farms consistent with the
county-by-farm-size counts (either provided for or estimated) and
(ii)~assigning farms to cells. The \textsc{GenFarms} algorithm for generating farms
is described in the supplement. 

The objective function encodes several minimization criteria. They are 
stated in order of priority: (i)~feasibility: modifies the subtype head 
count minimally to ensure feasibility of the assignment,  (ii)~equitable 
distribution of head counts for each subtype of livestock across farms within a
category, (iii)~minimize the number of subtypes within a farm, and (iv)~align with
known county-by-farm-size counts.
The  \textsc{FarmsToCells} algorithm (described in the supplement) assigns a cell to each farm. The objective of this algorithm is to ensure that
the head counts resulting from the assignment and \glw{} head counts are as
closely aligned as possible.

\subsection{Wild Bird Abundance and Movement}
We leveraged data from eBird's Status and Trends
products~\cite{eBirdStatusTrends2022} (see \eBird{} in
Table~\ref{tab:data}) to construct this component.  We recall that this
component~$\birds\big(\species, \abundance(\cdot), \birdnet\big)$ captures
spatiotemporal abundance and movement of multiple species of birds.
The \eBird{} 
data provides weekly estimates of 
relative bird abundance across a high-resolution grid~(2.96km$\times$2.96km).
These estimates, derived from ensemble machine learning models, represent 
the expected count of a species on a standardized eBird checklist at a 
given location and date. The models combine millions of citizen science
observations with
environmental predictors, accounting for factors such as land cover,
climate variables, and observation effort. A relative abundance of 1.0 for
a species at a particular location and time would indicate that an average
eBird checklist at that place and time would be expected to count one
individual of that species. Higher values indicate more individuals would
be expected, while lower values indicate the species would be observed less
frequently or in smaller numbers.  This approach
ensures that our model captures the most relevant species for studying
avian influenza transmission. We chose bird species for which H5N1 cases were
observed in the period 2022-2024. A total of~40 species of birds were identified, of 
which abundance data was available for~36 species.
For each of the selected species, we extract relative abundance values
along with their associated geographic coordinates for all 52 weeks in the
year. Our processing pipeline is described in the supplement.

\subsection{Dairy, Meat, and Egg Processing Plants}
We provide a layer for animal product processing plants with attributes such
as size, type of processing (dairy, meat, egg, etc.), livestock type, etc.
(see Table~\ref{tab:data}).  We use data from Agricultural Marketing
Services~\cite{usdadairy} for a list of large dairy processing plants.
These facilities process diverse categories of dairy products, ranging from
fluid milk and cream to various types of cheese, butter, and specialty
products identified by  product codes. We have developed a classification
of dairy plant codes based on the likelihood of handling unpasteurized
milk. This likelihood can help in spillover risk assessment as well as
inform surveillance strategies to detect HPAI incidences (through, for
example, bulk testing) in associated dairy farms. This classification is
presented in a table in the supplement that categorizes product codes into
high, medium, and low-risk groups.  We extract data from the Food
Safety and Inspection Service of the USDA which maintains a comprehensive
list of meat and egg processing establishments that need permits to be
operational.  For these plants, attributes such as size, address, plant
type (poultry or slaughter), and coordinates are available.  Data from both
sources were combined and standardized to form this layer.

\subsection{Human Population}
We develop a gridded representation of the US population with rich
demographic and employment-related attributes. This data is derived from a
synthetic population~\cite{chen2025epihiper,harrison2023synthetic,eubank2004modelling}
that is developed using diverse datasets such as census data, land use
data, activity patterns, building maps, etc. Each individual in the
population is associated with a residential location, an occupation identified by the Standard Occupational
Classification code (SOCP), and an industry, identified by the North American
Industry Classification System (NAICS) code. We identified all occupational
and industry codes that include livestock employment; these are listed in
the supplement. Individuals whose SOCP or NAICS codes did not belong to
this list were assigned a default code~$0$. For each combination of
demographic attributes~$\delta$ and employment~$\employ$, the population is
aggregated at the grid cell level. Our estimates of livestock worker counts are
consistent with the American Community Survey when aggregated to the 
US Census Public Use Microdata Area (PUMA) level.

\section{Usage Notes}
\label{sec:usage}
\subsection{Visualizations and Access}
The synthetic spatiotemporal dataset of interacting livestock and wild bird populations is designed to be easily accessible to and usable by researchers, policymakers, and modelers interested in studying avian influenza dynamics. 
We provide an interactive visualization dashboard, Digital Twin for Transboundary OneHealth (DiTTO) (shown in Figure~\ref{fig:dashboard}) to make the data available. 

The dashboard user interface is divided into three sections: a navigation bar, where users can indicate which data layers they are interested in viewing by population type and relevant subtypes (under Heatmap Measure), and even to pinpoint specific regions. On the lower left side of the screen is a heatmap where users can view where the selected population type is prevalent; users can view that data at US state resolutions, or click on the map to view county resolution heatmaps for the selected state. On the lower right side of the user interface is a data table where users can view the actual counts across all of the subtypes for the selected population type and region(s). Users can download the datasets in two ways from the web portal: (i) they can click on the Download Table button above the data table to download the queried rows displayed in the data table, or they can click on the Download Layer button on the map to download the complete grid-level layer data for the selected population type.

In short, DiTTO provides the following key features to make the datasets accessible to its users. 
\begin{itemize}
    \item Interactive maps showing the distribution of livestock, farms, wild bird populations, human populations, and processing centers at the state and county levels.
    \item The ability to search for specific regions for easier comparison.
    \item Filters for selecting specific regions, time periods (for wild birds), and livestock types.
    \item Download functionality for either the complete layer or for a subselection of that layer.
\end{itemize}


\begin{figure}[htb]
\centering
\includegraphics[width=\textwidth]{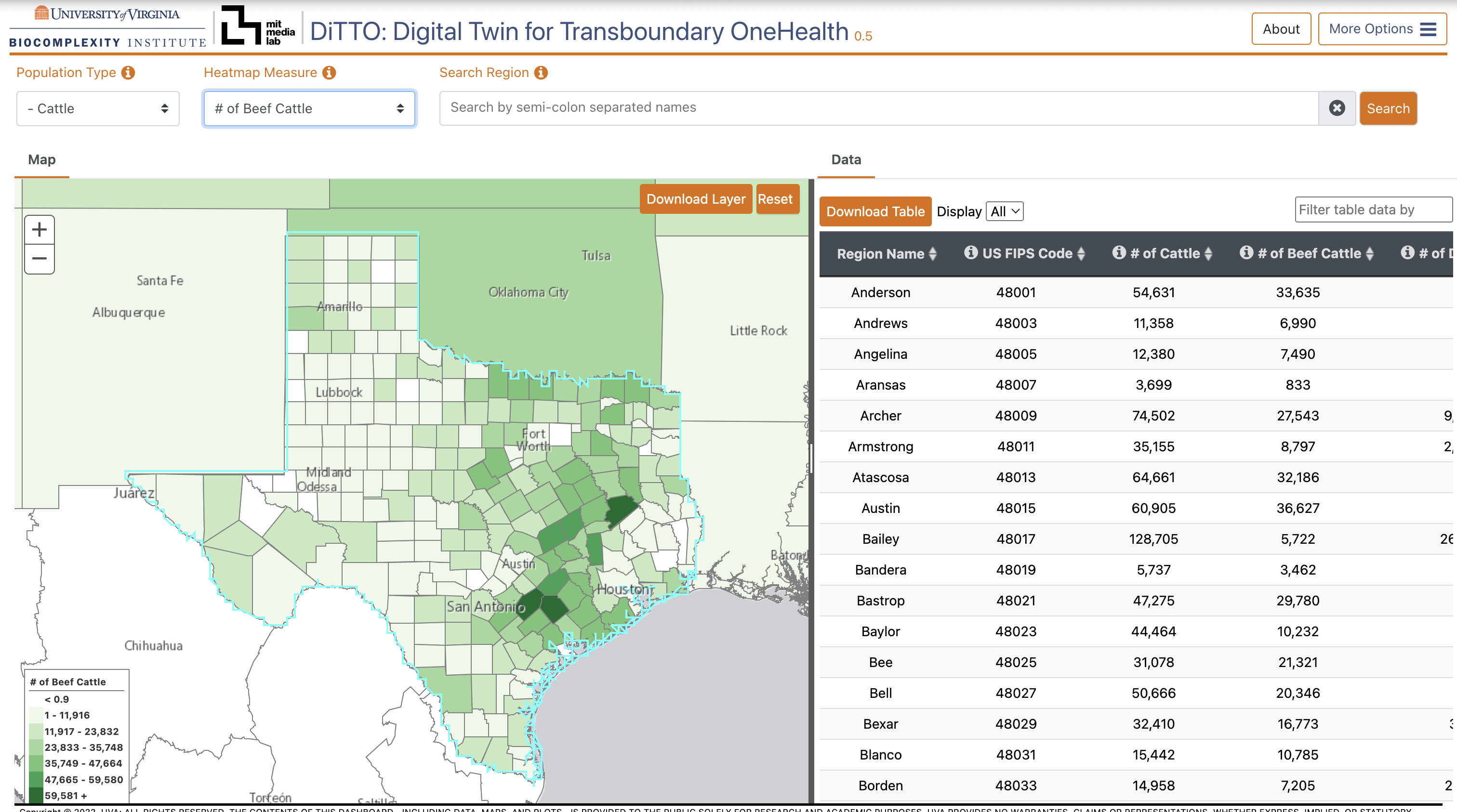}
\caption{The interactive visualization dashboard allows users to explore the livestock and avian populations in an interactive, spatiotemporal way. It is available at \url{https://ditto.bii.virginia.edu}.
}
\label{fig:dashboard}
\end{figure}

}
\bibliographystyle{plain}
\bibliography{refs}

\begin{thebibliography}{10}

\bibitem{abueg2020modeling}
Matthew Abueg, Robert Hinch, Neo Wu, Luyang Liu, William Probert, Austin Wu,
  Paul Eastham, Yusef Shafi, Matt Rosencrantz, Michael Dikovsky, et~al.
\newblock {M}odeling the {C}ombined {E}ffect of {D}igital {E}xposure
  {N}otification and {N}on-{P}harmaceutical {I}nterventions on the {COVID}-19
  {E}pidemic in {W}ashington {S}tate.
\newblock {\em MedRxiv}, pages 2020--08, 2020.

\bibitem{adiga2015generating}
Abhijin Adiga, Aditya Agashe, Shaikh Arifuzzaman, Christopher~L Barrett,
  Richard~J Beckman, Keith~R Bisset, Jiangzhuo Chen, Youngyun Chungbaek,
  Stephen~G Eubank, Sandeep Gupta, et~al.
\newblock Generating a synthetic population of the united states.
\newblock {\em Network Dynamics and Simulation Science Laboratory, Tech. Rep.
  NDSSL}, pages 15--009, 2015.

\bibitem{aleta2020modelling}
Alberto Aleta, David Martin-Corral, Ana Pastore~y Piontti, Marco Ajelli, Maria
  Litvinova, Matteo Chinazzi, Natalie~E Dean, M~Elizabeth Halloran, Ira~M
  Longini~Jr, Stefano Merler, et~al.
\newblock {M}odelling the {I}mpact of {T}esting, {C}ontact {T}racing and
  {H}ousehold {Q}uarantine on {S}econd {W}aves of {COVID}-19.
\newblock {\em Nature Human Behaviour}, 4(9):964--971, September 2020.

\bibitem{usdadairy}
USDA AMS.
\newblock Dairy plants surveyed and approved for usda grading service.
\newblock \url{https://apps.ams.usda.gov/dairy/ApprovedPlantList/}, 2024.
\newblock [Accessed 08-2024].

\bibitem{ush5n1birds}
APHIS.
\newblock Detections of highly pathogenic avian influenza in wild birds.
\newblock
  \url{https://www.aphis.usda.gov/livestock-poultry-disease/avian/avian-influenza/hpai-detections/wild-birds}.
\newblock [Accessed 01-2025].

\bibitem{ush5n1poultry}
USDA APHIS.
\newblock {C}onfirmations of {H}ighly {P}athogenic {A}vian {I}nfluenza in
  {C}ommercial and {B}ackyard {F}locks.
\newblock \url{https://www.aphis.usda.gov/livestock-poultry
  disease/avian/avian-influenza/hpai-detections/commercial backyard-flocks}.
\newblock [Accessed 01-2025].

\bibitem{barrett2013planning}
Christopher Barrett, Keith Bisset, Shridhar Chandan, Jiangzhuo Chen, Youngyun
  Chungbaek, Stephen Eubank, Yaman Evrenoso{\u{g}}lu, Bryan Lewis, Kristian
  Lum, Achla Marathe, et~al.
\newblock Planning and response in the aftermath of a large crisis: {A}n
  agent-based informatics framework.
\newblock In {\em 2013 Winter Simulations Conference (WSC)}, pages 1515--1526.
  IEEE, 2013.

\bibitem{batty2024digital}
Michael Batty.
\newblock Digital twins in city planning.
\newblock {\em Nature Computational Science}, 4(3):192--199, 2024.

\bibitem{bruhn2012synthesized}
Mark~C Bruhn, Breda Munoz, James Cajka, Gary Smith, Ross~J Curry, Diane~K
  Wagener, and William~D Wheaton.
\newblock Synthesized population databases: a geospatial database of us poultry
  farms.
\newblock {\em Methods report (RTI Press)}, page~1, 2012.

\bibitem{burdett2015simulating}
Christopher~L Burdett, Brian~R Kraus, Sarah~J Garza, Ryan~S Miller, and Kathe~E
  Bjork.
\newblock Simulating the distribution of individual livestock farms and their
  populations in the {U}nited {S}tates: {A}n example using domestic swine
  ({S}us scrofa domesticus) farms.
\newblock {\em PloS one}, 10(11):e0140338, 2015.

\bibitem{bls}
{B}ureau of~{L}abor {S}tatistics.
\newblock {Quarterly Census of Employment and Wages}.
\newblock \url{https://www.bls.gov/cew/}, 2023.
\newblock [Accessed 2025-01].

\bibitem{burrough2024highly}
Eric~R Burrough, Drew~R Magstadt, Barbara Petersen, Simon~J Timmermans,
  Phillip~C Gauger, Jianqiang Zhang, Chris Siepker, Marta Mainenti, Ganwu Li,
  Alexis~C Thompson, et~al.
\newblock Highly pathogenic avian influenza {A (H5N1)} clade 2.3. 4.4 b virus
  infection in domestic dairy cattle and cats, {U}nited {S}tates, 2024.
\newblock {\em Emerging infectious diseases}, 30(7):1335, 2024.

\bibitem{caldarelli2023role}
Guido Caldarelli, Elsa Arcaute, Marc Barthelemy, Michael Batty, Carlos
  Gershenson, Dirk Helbing, Stefano Mancuso, Yamir Moreno, Jos{\'e}~J Ramasco,
  C{\'e}line Rozenblat, et~al.
\newblock The role of complexity for digital twins of cities.
\newblock {\em Nature Computational Science}, 3(5):374--381, 2023.

\bibitem{caliendo2022transatlantic}
Valentina Caliendo, NS~Lewis, A~Pohlmann, SR~Baillie, AC~Banyard, Martin Beer,
  IH~Brown, RAM Fouchier, RDE Hansen, TK~Lameris, et~al.
\newblock Transatlantic spread of highly pathogenic avian influenza {H5N1} by
  wild birds from {E}urope to {N}orth {A}merica in 2021.
\newblock {\em Scientific reports}, 12(1):11729, 2022.

\bibitem{caserta2024spillover}
Leonardo~C Caserta, Elisha~A Frye, Salman~L Butt, Melissa Laverack, Mohammed
  Nooruzzaman, Lina~M Covaleda, Alexis~C Thompson, Melanie~Prarat Koscielny,
  Brittany Cronk, Ashley Johnson, et~al.
\newblock Spillover of highly pathogenic avian influenza h5n1 virus to dairy
  cattle.
\newblock {\em Nature}, pages 1--8, 2024.

\bibitem{ush5n1humans}
CDC.
\newblock H5 bird flu: Current situation.
\newblock \url{https://www.cdc.gov/bird-flu/situation-summary/index.html}.
\newblock [Accessed 01-2025].

\bibitem{chen2022effective}
Jiangzhuo Chen, Stefan Hoops, Achla Marathe, Henning Mortveit, Bryan Lewis,
  Srinivasan Venkatramanan, Arash Haddadan, Parantapa Bhattacharya, Abhijin
  Adiga, Anil Vullikanti, et~al.
\newblock Effective social network-based allocation of covid-19 vaccines.
\newblock In {\em Proceedings of the 28th ACM SIGKDD Conference on Knowledge
  Discovery and Data Mining}, pages 4675--4683, 2022.

\bibitem{chen2025epihiper}
Jiangzhuo Chen, Stefan Hoops, Henning~S Mortveit, Bryan~L Lewis, Dustin Machi,
  Parantapa Bhattacharya, Srinivasan Venkatramanan, Mandy~L Wilson, Chris~L
  Barrett, and Madhav~V Marathe.
\newblock Epihiper—a high performance computational modeling framework to
  support epidemic science.
\newblock {\em PNAS nexus}, 4(1):pgae557, 2025.

\bibitem{cheng2023maps}
Mingjin Cheng, Xin Liu, Hu~Sheng, and Zengwei Yuan.
\newblock {MAPS}: {A} new model using data fusion to enhance the accuracy of
  high-resolution mapping for livestock production systems.
\newblock {\em One Earth}, 6(9):1190--1201, 2023.

\bibitem{delgado2019big}
Jorge~A Delgado, Nicholas~M Short~Jr, Daniel~P Roberts, and Bruce Vandenberg.
\newblock Big data analysis for sustainable agriculture on a geospatial cloud
  framework.
\newblock {\em Frontiers in Sustainable Food Systems}, 3:54, 2019.

\bibitem{deming1940least}
W.~Edwards Deming and Frederick~F. Stephan.
\newblock On a least squares adjustment of a sampled frequency table when the
  expected marginal totals are known.
\newblock {\em The Annals of Mathematical Statistics}, 11(4):427--444, 1940.

\bibitem{agcensus_layer}
ESRI.
\newblock {USDA} {C}ensus of {A}griculture 2017 - cattle production.
\newblock
  \url{https://www.arcgis.com/home/item.html?id=53137233a760432bb07c417eb3d758b8}.
\newblock [Accessed 06-03-2024].

\bibitem{eubank2004modelling}
Stephen Eubank, Hasan Guclu, VS~Anil~Kumar, Madhav~V Marathe, Aravind
  Srinivasan, Zoltan Toroczkai, and Nan Wang.
\newblock {M}odelling {D}isease {O}utbreaks in {R}ealistic {U}rban {S}ocial
  {N}etworks.
\newblock {\em Nature}, 429(6988):180--184, 2004.

\bibitem{faoGriddedLivestock}
FAO.
\newblock {G}{L}{W} 4: {G}ridded {L}ivestock {D}ensity.
\newblock
  \url{https://data.apps.fao.org/catalog/dataset/15f8c56c-5499-45d5-bd89-59ef6c026704}.
\newblock [Accessed 06-03-2024].

\bibitem{ferretti2020quantifying}
Luca Ferretti, Chris Wymant, Michelle Kendall, Lele Zhao, Anel Nurtay, Lucie
  Abeler-D{\"o}rner, Michael Parker, David Bonsall, and Christophe Fraser.
\newblock {Q}uantifying {SARS}-{CoV}-2 {T}ransmission {S}uggests {E}pidemic
  {C}ontrol with {D}igital {C}ontact {T}racing.
\newblock {\em Science}, 368(6491):eabb6936, 2020.

\bibitem{fienberg1970iterative}
Stephen~E. Fienberg.
\newblock An iterative procedure for estimation in contingency tables.
\newblock {\em The Annals of Mathematical Statistics}, 41(3):907--917, 1970.

\bibitem{eBirdStatusTrends2022}
Daniel Fink, Tom Auer, Alison Johnston, Matt Strimas-Mackey, Shawn Ligocki,
  Orin Robinson, Wesley Hochachka, Lauren Jaromczyk, Cynthia Crowlye, Kylee
  Dunham, Andrew Stillman, Ian Davies, Amanda Rodewald, Viviana Ruiz-Gutierrez,
  and Chris Wood.
\newblock e{B}ird status and trends, data version: 2022; released: 2023, 2023.

\bibitem{usdameat}
USDA FSIS.
\newblock Meat, poultry and egg product inspection directory.
\newblock
  \url{https://www.fsis.usda.gov/inspection/establishments/meat-poultry-and-egg-product-inspection-directory},
  2024.
\newblock [Accessed 08-2024].

\bibitem{fuentes2023birdflow}
Miguel Fuentes, Benjamin~M Van~Doren, Daniel Fink, and Daniel Sheldon.
\newblock Birdflow: Learning seasonal bird movements from ebird data.
\newblock {\em Methods in Ecology and Evolution}, 14(3):923--938, 2023.

\bibitem{gilbert2018global}
Marius Gilbert, Ga{\"e}lle Nicolas, Giusepina Cinardi, Thomas~P Van~Boeckel,
  Sophie~O Vanwambeke, GR~Wint, and Timothy~P Robinson.
\newblock Global distribution data for cattle, buffaloes, horses, sheep, goats,
  pigs, chickens and ducks in 2010.
\newblock {\em Scientific data}, 5(1):1--11, 2018.

\bibitem{handan2019deep}
Cassandra Handan-Nader and Daniel~E Ho.
\newblock Deep learning to map concentrated animal feeding operations.
\newblock {\em Nature Sustainability}, 2(4):298--306, 2019.

\bibitem{harrison2023synthetic}
Galen Harrison, Przemyslaw Porebski, Jiangzhuo Chen, Mandy Wilson, Henning
  Mortveit, Parantapa Bhattacharya, Dawen Xie, Stefan Hoops, Anil Vullikanti,
  Li~Xiong, et~al.
\newblock Synthetic information and digital twins for pandemic science:
  Challenges and opportunities.
\newblock In {\em 2023 5th IEEE International Conference on Trust, Privacy and
  Security in Intelligent Systems and Applications (TPS-ISA)}, pages 23--33.
  IEEE, 2023.

\bibitem{hoops2021high}
Stefan Hoops, Jiangzhuo Chen, Abhijin Adiga, Bryan Lewis, Henning Mortveit,
  Hannah Baek, Mandy Wilson, Dawen Xie, Samarth Swarup, Srinivasan
  Venkatramanan, et~al.
\newblock {H}igh {P}erformance {A}gent-{B}ased {M}odeling to {S}tudy
  {R}ealistic {C}ontact {T}racing {P}rotocols.
\newblock In Kim Sojung, Ben Feng, Katy Smith, Sara Masoud, and Zeyu Zheng,
  editors, {\em 2021 Winter Simulation Conference (WSC)}, pages 1--12. IEEE,
  2021.

\bibitem{humphreys2020waterfowl}
John~M Humphreys, Andrew~M Ramey, David~C Douglas, Jennifer~M Mullinax,
  Catherine Soos, Paul Link, Patrick Walther, and Diann~J Prosser.
\newblock Waterfowl occurrence and residence time as indicators of h5 and h7
  avian influenza in north american poultry.
\newblock {\em Scientific Reports}, 10(1):2592, 2020.

\bibitem{kent2022spatiotemporal}
Cody~M Kent, Andrew~M Ramey, Joshua~T Ackerman, Justin Bahl, Sarah~N Bevins,
  Andrew~S Bowman, Walter~M Boyce, Carol~J Cardona, Michael~L Casazza, Troy~D
  Cline, et~al.
\newblock Spatiotemporal changes in influenza a virus prevalence among wild
  waterfowl inhabiting the continental united states throughout the annual
  cycle.
\newblock {\em Scientific reports}, 12(1):13083, 2022.

\bibitem{kerr2021covasim}
Cliff~C Kerr, Robyn~M Stuart, Dina Mistry, Romesh~G Abeysuriya, Katherine
  Rosenfeld, Gregory~R Hart, Rafael~C N{\'u}{\~n}ez, Jamie~A Cohen, Prashanth
  Selvaraj, Brittany Hagedorn, et~al.
\newblock {C}ovasim: {A}n {A}gent-{B}ased {M}odel of {COVID}-19 {D}ynamics and
  {I}nterventions.
\newblock {\em PLOS Computational Biology}, 17(7):e1009149, 2021.

\bibitem{koopmans2024panzootic}
Marion~PG Koopmans, Casey~Barton Behravesh, Andrew~A Cunningham, Wiku~B
  Adisasmito, Salama Almuhairi, P{\'e}p{\'e} Bilivogui, Salome~A Bukachi,
  Natalia Casas, Natalia~Cediel Becerra, Dominique~F Charron, et~al.
\newblock The panzootic spread of highly pathogenic avian influenza h5n1
  sublineage 2.3. 4.4 b: a critical appraisal of one health preparedness and
  prevention.
\newblock {\em The Lancet Infectious Diseases}, 24(12):e774--e781, 2024.

\bibitem{leguia2023highly}
Mariana Leguia, Alejandra Garcia-Glaessner, Breno Mu{\~n}oz-Saavedra, Diana
  Juarez, Patricia Barrera, Carlos Calvo-Mac, Javier Jara, Walter Silva, Karl
  Ploog, Lady Amaro, et~al.
\newblock Highly pathogenic avian influenza {A (H5N1)} in marine mammals and
  seabirds in {P}eru.
\newblock {\em Nature Communications}, 14(1):5489, 2023.

\bibitem{livestockzoonoses}
Kacper Libera, Kacper Konieczny, Julia Grabska, Wiktoria Szopka, Agata
  Augustyniak, and Ma{\l}gorzata Pomorska-M{\'o}l.
\newblock Selected livestock-associated zoonoses as a growing challenge for
  public health.
\newblock {\em Infectious disease reports}, 14(1):63--81, 2022.

\bibitem{lloyd2017high}
Christopher~T Lloyd, Alessandro Sorichetta, and Andrew~J Tatem.
\newblock High resolution global gridded data for use in population studies.
\newblock {\em Scientific data}, 4(1):1--17, 2017.

\bibitem{macdonald2009transformation}
James~M MacDonald and William~D McBride.
\newblock The transformation of us livestock agriculture scale, efficiency, and
  risks.
\newblock Economic Information Bulletin No. 43, Economic Research Service, U.S.
  Dept. of Agriculture, 2009.

\bibitem{marathe2014prescriptive}
Madhav~V Marathe, Henning~S Mortveit, Nidhi Parikh, and Samarth Swarup.
\newblock Prescriptive analytics using synthetic information.
\newblock In {\em Emerging Methods in Predictive Analytics: Risk Management and
  Decision-Making}, pages 1--19. IGI Global, 2014.

\bibitem{meyur2022ensembles}
Rounak Meyur, Anil Vullikanti, Samarth Swarup, Henning~S Mortveit, Virgilio
  Centeno, Arun Phadke, H~Vincent Poor, and Madhav~V Marathe.
\newblock Ensembles of realistic power distribution networks.
\newblock {\em Proceedings of the National Academy of Sciences},
  119(42):e2205772119, 2022.

\bibitem{mihai2022digital}
Stefan Mihai, Mahnoor Yaqoob, Dang~V Hung, William Davis, Praveer Towakel,
  Mohsin Raza, Mehmet Karamanoglu, Balbir Barn, Dattaprasad Shetve, Raja~V
  Prasad, et~al.
\newblock Digital twins: A survey on enabling technologies, challenges, trends
  and future prospects.
\newblock {\em IEEE Communications Surveys \& Tutorials}, 24(4):2255--2291,
  2022.

\bibitem{westnile2019spatiotemporal}
Mark~H Myer and John~M Johnston.
\newblock Spatiotemporal bayesian modeling of west nile virus: Identifying risk
  of infection in mosquitoes with local-scale predictors.
\newblock {\em Science of the Total Environment}, 650:2818--2829, 2019.

\bibitem{agcensus2022}
{U}{S}{D}{A} {N}ational {A}gricultural~{S}tatistics {S}ervice.
\newblock {C}ensus of {A}griculture.
\newblock \url{https://www.nass.usda.gov/AgCensus/}.
\newblock [Accessed 03-Jan-2023].

\bibitem{agcensus2022download}
{U}{S}{D}{A} {N}ational {A}gricultural~{S}tatistics {S}ervice.
\newblock {C}ensus of {A}griculture.
\newblock \url{https://www.nass.usda.gov/datasets/qs.census2022.txt.gz.}
\newblock [Accessed 11-Jun-2024].

\bibitem{nguyen2024emergence}
Thao-Quyen Nguyen, Carl Hutter, Alexey Markin, Megan~N Thomas, Kristina Lantz,
  Mary~Lea Killian, Garrett~M Janzen, Sriram Vijendran, Sanket Wagle, Blake
  Inderski, et~al.
\newblock Emergence and interstate spread of highly pathogenic avian influenza
  a (h5n1) in dairy cattle.
\newblock {\em bioRxiv}, pages 2024--05, 2024.

\bibitem{cafomaps}
The~University of~Iowa.
\newblock {C}{A}{F}{O}s in the {U}{S}.
\newblock \url{https://cafomaps.org/}.
\newblock [Accessed 08-19-2024].

\bibitem{paim2019epidemiology}
Francine~C Paim, Andrew~S Bowman, Lauren Miller, Brandi~J Feehan, Douglas
  Marthaler, Linda~J Saif, and Anastasia~N Vlasova.
\newblock Epidemiology of deltacoronaviruses ($\delta$-cov) and
  gammacoronaviruses ($\gamma$-cov) in wild birds in the united states.
\newblock {\em Viruses}, 11(10):897, 2019.

\bibitem{patyk2020modelling}
Kelly~A Patyk, Mary~J McCool-Eye, David~D South, Christopher~L Burdett, Susan~A
  Maroney, Andrew Fox, Grace Kuiper, and Sheryl Magzamen.
\newblock Modelling the domestic poultry population in the united states: A
  novel approach leveraging remote sensing and synthetic data methods.
\newblock {\em Geospatial Health}, 15(2), 2020.

\bibitem{statecafoguides}
{SRAP} project.
\newblock State {CAFO} guides.
\newblock \url{https://sraproject.org/state-cafo-guides/}.
\newblock [Accessed 10-22-2024].

\bibitem{prosser2024using}
Diann~J Prosser, Cody~M Kent, Jeffery~D Sullivan, Kelly~A Patyk, Mary-Jane
  McCool, Mia~Kim Torchetti, Kristina Lantz, and Jennifer~M Mullinax.
\newblock Using an adaptive modeling framework to identify avian influenza
  spillover risk at the wild-domestic interface.
\newblock {\em Scientific Reports}, 14(1):14199, 2024.

\bibitem{puryear2023highly}
Wendy Puryear, Kaitlin Sawatzki, Nichola Hill, Alexa Foss, Jonathon~J Stone,
  Lynda Doughty, Dominique Walk, Katie Gilbert, Maureen Murray, Elena Cox,
  et~al.
\newblock Highly pathogenic avian influenza {A (H5N1)} virus outbreak in {N}ew
  {E}ngland seals, {U}nited {S}tates.
\newblock {\em Emerging Infectious Diseases}, 29(4):786, 2023.

\bibitem{pylianidis2021introducing}
Christos Pylianidis, Sjoukje Osinga, and Ioannis~N Athanasiadis.
\newblock Introducing digital twins to agriculture.
\newblock {\em Computers and Electronics in Agriculture}, 184:105942, 2021.

\bibitem{robinson2022mapping}
Caleb Robinson, Ben Chugg, Brandon Anderson, Juan M~Lavista Ferres, and
  Daniel~E Ho.
\newblock Mapping industrial poultry operations at scale with deep learning and
  aerial imagery.
\newblock {\em IEEE Journal of Selected Topics in Applied Earth Observations
  and Remote Sensing}, 15:7458--7471, 2022.

\bibitem{robinson2014mapping}
Timothy~P Robinson, GR~William Wint, Giulia Conchedda, Thomas~P Van~Boeckel,
  Valentina Ercoli, Elisa Palamara, Giuseppina Cinardi, Laura D'Aietti, Simon~I
  Hay, and Marius Gilbert.
\newblock Mapping the global distribution of livestock.
\newblock {\em PloS one}, 9(5):e96084, 2014.

\bibitem{sullivan2009ebird}
Brian~L Sullivan, Christopher~L Wood, Marshall~J Iliff, Rick~E Bonney, Daniel
  Fink, and Steve Kelling.
\newblock e{B}ird: {A} citizen-based bird observation network in the biological
  sciences.
\newblock {\em Biological conservation}, 142(10):2282--2292, 2009.

\bibitem{thorve2023high}
Swapna Thorve, Young~Yun Baek, Samarth Swarup, Henning Mortveit, Achla Marathe,
  Anil Vullikanti, and Madhav Marathe.
\newblock High resolution synthetic residential energy use profiles for the
  {U}nited {S}tates.
\newblock {\em Scientific Data}, 10(1):76, 2023.

\bibitem{uhart2024massive}
Marcela~M Uhart, Ralph~ET Vanstreels, Martha~I Nelson, Valeria Olivera, Julieta
  Campagna, Victoria Zavattieri, Philippe Lemey, Claudio Campagna, Valeria
  Falabella, and Agustina Rimondi.
\newblock Massive outbreak of {I}nfluenza {A H5N1} in elephant seals at
  {P}eninsula {V}aldes, {A}rgentina: increased evidence for mammal-to-mammal
  transmission.
\newblock {\em bioRxiv}, pages 2024--05, 2024.

\bibitem{van2021challenges}
Mary van Andel, Michael~J Tildesley, and M~Carolyn Gates.
\newblock Challenges and opportunities for using national animal datasets to
  support foot-and-mouth disease control.
\newblock {\em Transboundary and Emerging Diseases}, 68(4):1800--1813, 2021.

\bibitem{woahh5n1}
{World Organization for Animal Health}.
\newblock {United States of America - Influenza A viruses of high
  pathogenicity}.
\newblock \url{https://wahis.woah.org/#/in-event/4451/dashboard}.
\newblock [Accessed 01-2025].

\bibitem{wu2021digital}
Yiwen Wu, Ke~Zhang, and Yan Zhang.
\newblock Digital twin networks: A survey.
\newblock {\em IEEE Internet of Things Journal}, 8(18):13789--13804, 2021.

\bibitem{xiao2006semi}
Yanni Xiao, Damian Clancy, Nigel~P French, and Roger~G Bowers.
\newblock A semi-stochastic model for salmonella infection in a multi-group
  herd.
\newblock {\em Mathematical Biosciences}, 200(2):214--233, 2006.

\bibitem{yuan2023synthetic}
Rui Yuan, S~Ali Pourmousavi, Wen~L Soong, Andrew~J Black, Jon~AR Liisberg, and
  Julian Lemos-Vinasco.
\newblock A synthetic dataset of {D}anish residential electricity prosumers.
\newblock {\em Scientific Data}, 10(1):371, 2023.

\bibitem{zhu2024high}
Chuanyong Zhu, Renqiang Li, Mengyi Qiu, Changtong Zhu, Yichao Gai, Ling Li,
  Na~Yang, Lei Sun, Chen Wang, Baolin Wang, et~al.
\newblock High spatiotemporal resolution ammonia emission inventory from
  typical industrial and agricultural province of {C}hina from 2000 to 2020.
\newblock {\em Science of The Total Environment}, 918:170732, 2024.

\end{thebibliography}
\iftoggle{abstract}{
}{

\renewcommand{\baselinestretch}{0.9}
\baselineskip=\normalbaselineskip

\clearpage

\appendix

\baselineskip=\normalbaselineskip

\setcounter{figure}{0}
\renewcommand{\thefigure}{S\arabic{figure}}

\begin{center}
\fbox{{\Large\textbf{Supplementary Information}}}
\end{center}

\bigskip

\section{Livestock}
\label{ssec:livestock}
\subsection{Organization and Preliminary Definitions}
The types of livestock covered in this work are shown in
Table~\ref{tab:agcensus}.  Two data sources were used to construct the
livestock layers: Census of Agriculture (\agcensus{}) and Gridded Livestock
of the World (\glw{}). These are described in the following sections. The
overview of the methodology used to fill gaps, generate farms, and
assign farms to grid cells is captured in Figure~\ref{fig:livestock}. The
description of the same appears in Sections~\ref{sec:fill_gaps}
and~\ref{sse:farms_to_cells}.

A \textbf{livestock type} refers to a class of animals.  Examples of
livestock types include cattle, poultry, hogs, and sheep. A \textbf{livestock
subtype} represents a subclass of animals within a livestock type.  For
example, the livestock type cattle includes subtypes such as beef cows and
milk cows.  Likewise, the livestock type poultry includes subtypes such as
(egg)~layers, pullets, turkeys, etc.

\newcommand{\Hi}{H_i}
\newcommand{\Fi}{F_i}
\newcommand{\Hg}{H_{\gamma}}
\newcommand{\Fg}{F_{\gamma}}
\newcommand{\Hig}{H_{i\gamma}}
\newcommand{\Fig}{F_{i\gamma}}
\newcommand{\hig}{h_{i\gamma}}
\newcommand{\hifg}{h_{if\gamma}}
\newcommand{\xifgk}{x_{if\gamma k}}
\newcommand{\yifgk}{y_{if\gamma k}}
\newcommand{\zifgk}{z_{if\gamma k}}
\newcommand{\wkmin}{\mbox{$W_k^{\mathrm{min}}$}}
\newcommand{\wkmax}{\mbox{$W_k^{\mathrm{max}}$}}
\newcommand{\Hgk}{H_{\gamma k}}
\newcommand{\Fgk}{F_{\gamma k}}

\subsection{Census of Agriculture (\agcensus)}
\label{sec:agcensus}
\subsubsection{Data organization and availability}
\agcensus{} provides counts of heads (i.e., population size) and 
farms for various livestock types and subtypes. The data is available at
three different administrative levels -- country, state, and county. The
livestock types we consider here are cattle, poultry, hogs, and sheep. We also consider 
various subtypes for cattle and poultry. These types and subtypes are summarized in
Table~\ref{tab:agcensus}. The data organization is livestock type specific,
making it a non-trivial task to extract relevant
information. For each administrative level, the total counts are provided.
Also provided are counts corresponding to different farm sizes. Farms are
binned into categories based on the head counts of the corresponding livestock type, such as 1–24, 25–49, 50–99, 100–199, 200–499,
500–999, and 1000 or more, where each category is specified by the minimum
and maximum head count, respectively, in the member farm. 
Accordingly, we
have four types of counts: (i)~state-total, (ii)~state-by-farm-size,
(iii)~county-total, and (iv)~county-by-farm-size. Table~\ref{tab:ftoc_input}
depicts this organization of counts at the county level. The set of farm
categories for each subtype is consistent with that of its parent livestock
type.

\begin{table}[tb]
\centering
\caption{Population and operations statistics for various livestock types covered by our synthetic dataset. Also shown are the counts after filling gaps.}
\small
\begin{tabular}{llrrrrrr}
\toprule
 &  & \multicolumn{4}{r}{heads} & \multicolumn{2}{r}{farms} \\
 &  & state tot. & county tot. & filled gaps & final & state & processed \\
livestock & subtype &  &  &  &  &  &  \\
\midrule
\multirow[t]{4}{*}{cattle} & all & 87954742 & 85973763 & 85973763 & 87932032 & 732123 & 731981 \\
 & beef & 29214479 & 27790671 & 29207376 & 29199243 & 622162 & 622050 \\
 & milk & 9309855 & 7526842 & 9317802 & 9317612 & 36024 & 35996 \\
 & other & 49430408 & 46255380 & 49422350 & 49415177 & 594222 & 594108 \\
\cline{1-8}
hogs & all & 73645928 & 62541219 & 73810004 & 73808393 & 60809 & 60731 \\
\cline{1-8}
\multirow[t]{18}{*}{poultry} & chukars & 1036946 & 621024 & 1048787 & 1047104 & 801 & 800 \\
 & ckn-broilers & 1737674957 & 1680674087 & 1737674725 & 1737795431 & 42991 & 42947 \\
 & ckn-layers & 375927945 & 199001866 & 388508984 & 389641754 & 240530 & 240270 \\
 & ckn-pullets & 139203843 & 82678506 & 144030350 & 144029544 & 34874 & 34829 \\
 & ckn-roosters & 7656478 & 6858454 & 7720552 & 7720400 & 42110 & 42064 \\
 & ducks & 4341317 & 3422540 & 4448858 & 4448287 & 34781 & 34724 \\
 & emus & 12538 & 9462 & 12440 & 12427 & 1566 & 1561 \\
 & geese & 101823 & 83174 & 101521 & 101320 & 11940 & 11911 \\
 & guineas & 391931 & 340504 & 391674 & 391549 & 18853 & 18844 \\
 & ostriches & 2245 & 1519 & 3496 & 3496 & 232 & 232 \\
 & partridges & 49162 & 9462 & 61147 & 61147 & 68 & 68 \\
 & peafowl & 54947 & 42795 & 54679 & 54669 & 6930 & 6928 \\
 & pheasants & 3187136 & 1243764 & 3279830 & 3266790 & 2257 & 2255 \\
 & pigeons & 212559 & 160312 & 302934 & 285600 & 2196 & 2194 \\
 & poultry-other & 69840 & 37923 & 84241 & 84241 & 789 & 789 \\
 & quail & 9188443 & 6245272 & 9294150 & 9293769 & 4738 & 4731 \\
 & rheas & 1013 & 382 & 1122 & 1122 & 152 & 152 \\
 & turkeys & 97064430 & 84529090 & 97312274 & 97311591 & 23431 & 23373 \\
\cline{1-8}
sheep & all & 5104328 & 3664088 & 5103716 & 5102574 & 88853 & 88795 \\
\cline{1-8}
\bottomrule
\end{tabular}

\label{tab:agcensus}
\end{table}

\paragraph{Missing data.}
In many instances, head counts are redacted. The more refined the count
category, the greater the incidence of missing data. There are more instances of missing data 
(i)~at the county level compared to state level, (ii)~in the farm-size
categories compared to total head counts, and (iii)~in subtype counts compared
to total livestock counts. This can be observed in Table~\ref{tab:agcensus}
for head counts aggregated using different types of counts. Operation counts, on the other hand, are always provided. For all livestock types except poultry, farm
counts are provided at every administrative level for every farm category.

\begin{table}[htb]
\caption{The data format for state-by-farm-size and county-by-farm-size. Some
of the head counts data is redacted. The corresponding totals (either state
or county) are denoted by $H$ (for ``all''), $H_\text{beef}$, $H_{milk}$
and $H_{other}$. Depending on the instance, any of the totals or counts by
farm size can be missing.}
\centering
\begin{tabular}{cc|cccc}
\bf Cat. & \bf Farm size & \bf all & \bf beef & \bf milk & \bf other\\
\midrule
1 & 1--9 & $(F_1,H_1)$ & $(F_{1,\text{beef}},H_{1,\text{beef}})$ &
$(F_{1,\text{milk}},H_{1,\text{milk}})$ &
$(F_{1,\text{other}},H_{1,\text{other}})$ \\
2 & 10--19 & $(F_2,H_2)$ & $(F_{2,\text{beef}},H_{2,\text{beef}})$ &
$(F_{2,\text{milk}},H_{2,\text{milk}})$ &
$(F_{2,\text{other}},H_{2,\text{other}})$ \\
 & $\vdots$ &  &  & $\cdots$ &  \\
$i$ & $\wimin$--$\wimax$ & $(\Fi, \Hi)$  & $(F_{i,\text{beef}},H_{i,\text{beef}})$ &
$(F_{i,\text{milk}},H_{i,\text{milk}})$ &
$(F_{i,\text{other}},H_{i,\text{other}})$  \\
 & $\vdots$ &  &  & $\cdots$ &  \\
\end{tabular}
\label{tab:ftoc_input}
\end{table}

\paragraph{Notation.} We set up some formal notation here to facilitate the
description of our framework. Since each livestock type is processed
independently, the notation will not carry information about the livestock
type; it is assumed that the livestock type is known. The same holds true for
the administrative level. Given a livestock type, the number of categories
is denoted by~$\ell$. Then, for~$i=1,\ldots,\ell$, a category is specified
by~$(\wimin,\wimax)$, the minimum and maximum population sizes
respectively. Given an administrative level, let~$H$ denote the total head
count and~$F$ denote the total farm count. For farm category~$i$, let $\Hi$
and~$\Fi$ denote the head and farm counts, respectively.
Let~$\Gamma$ denote the set of different subtypes. The notation is similar to the
one developed above; for a subtype~$\gamma\in\Gamma$,~$\Hg$ and~$\Fg$
denote the total counts of the subtype at the target administrative
level, and~$\Hgk$ and~$\Fgk$ denote the farm category specific counts for
category~$k=1,\ldots,\ell$. 

\subsubsection{Livestock type-specific information} 
The relevant head and farm counts were extracted from the full AgCensus dataset by querying out the rows where
\ttt{statisticcat\_desc="INVENTORY"}; these rows are further filtered by  \ttt{unit\_desc="HEAD"} or \ttt{unit\_desc="OPERATIONS"}, depending on whether head counts or number of operations are being calculated, respectively. The
state- and county-level counts were extracted by filtering
\ttt{agg\_level} to \ttt{STATE} and \ttt{COUNTY}, respectively.

\paragraph{Cattle.}
The counts were obtained by extracting rows where
\ttt{commodity\_desc="CATTLE"}. The cattle
population is categorized into three subtypes, \ttt{beef}, \ttt{milk}, and
\ttt{other} (specified by the \ttt{class\_desc} field). The total count of
cattle was obtained by setting \ttt{class\_desc="INCL CALVES"}. The state-
and county-total counts were obtained by extracting rows where
\ttt{domain\_desc="TOTAL"}. The state-by-farm-size and county-by-farm-size
counts were obtained by setting \ttt{domain\_desc$\ne$"TOTAL"}.  There are
several additional conditions that had to be filtered to get the
appropriate counts. These conditions only include rows where (i)
\ttt{domaincat\_desc} text includes text "0 HEAD" or "1 OR MORE HEAD"; (ii)
(\ttt{domain\_desc} includes text such as "inventory of milk/beef cows" or
"inventory of cows" and (iii) \ttt{class\_desc} is either "INCL CALVES" or
"EXCL COWS").

\paragraph{Poultry.}
Poultry data has many subtypes. In \agcensus{} each subtype is organized as
separate livestock under the group poultry, i.e., \ttt{group\_desc="POULTRY"}.
We remapped this data by creating a new livestock called \ttt{poultry} and mapping all livestock
under it to distinct subtypes.  The counts for chickens were obtained by
setting \ttt{commodity\_desc="CHICKENS"}. 
There are four subtypes corresponding to
chickens: layers (\ttt{ckn-layers}), broilers (\ttt{ckn-broilers}), pullets
(\ttt{ckn-pullets}), and roosters (\ttt{ckn-roosters}). For layers,
state-level counts of farms by category is present. However, we have
ignored this as the corresponding head counts are absent.
The counts of other poultry such as turkeys and ducks was obtained by
setting \ttt{group\_desc="POULTRY"} and \ttt{commodity\_desc$\ne$"CHICKENS"}.

\paragraph{Hogs.}
The counts were obtained by setting \ttt{commodity\_desc="HOGS"}.
The rest are similar to cattle.

\paragraph{Sheep.} The counts were obtained by setting
\ttt{commodity\_desc="SHEEP"} 
and \ttt{class="INCL LAMBS"}. No subtypes were
considered. The remaining details are similar to cattle.

\paragraph{Aligning farm categories.} Farm category specifications are more
refined at the state level than at the county level. For example, at the
county level, the largest category is 500 or more, whereas at the state
level, there are categories such as `1000-2499' and `2500 or more'. We map
all state-level categories to county-level categories by either aggregating
the counts in the additional categories or simply removing the
categories if they had already been accounted for at the county level.
\newcommand{\glwx}{\texttt{x}} 
\newcommand{\glwy}{\texttt{y}} 

\subsection{Gridded Livestock of the World}
\label{sec:glw}
The Gridded Livestock of the World~\cite{faoGriddedLivestock} dataset
provides a gridded distribution of livestock abundance at~$5$~arc minute
resolution.  The gridded distribution data was constructed by combining
detailed livestock census statistics mined from various sources using random
forest models with predictors of the following types: land use, human
population, travel times, vegetation, and climate. Unsuitable areas such
as water bodies and core urban centers are identified using land cover and
human population density information. More details are provided in
Gilbert~et~al.~\cite{gilbert2018global}.

The data is available in the following format. Each grid cell is identified
by a cell ID denoted by a pair of integers, (\glwx, \glwy). For each grid
cell, if a livestock abundance is available, the livestock type and value
are provided.  Among the livestock types or species provided, we considered
cattle, buffaloes, sheep, pigs, chickens, and ducks. Buffaloes were mapped
to cattle, and ducks and chickens were mapped to poultry. We did not
consider goats and horses. No information on livestock subtypes is
provided.

We identified the cells corresponding to the contiguous US and associated
them with their respective state and county FIPS codes. The cells are
denoted by~$C_j$ while the abundance value of a livestock type is denoted
by~$Q_j$. The notation does not include livestock type as each type is
processed independently.
\subsection{Filling gaps} 
\label{sec:fill_gaps}
As mentioned above, some head counts are missing in all count types:
state-total, state-by-farm-size, county-total, and county-by-farm-size.
We use a combination of integer linear programs~(ILPs) and iterative
proportional fitting~(IPF) following
Burdett~et~al.~\cite{burdett2015simulating} to address these omissions. For cattle and poultry, gaps
are filled for each subtype, while for hogs and sheep they are filled for
the livestock type.

We use the integer program Algorithm~\ref{alg:fillgap} to fill missing data
for the following types of counts: state-total, state-by-farm-size, and
county-total. It takes as input all the known counts, sum of all the
counts, and bounds on the unknown counts, and distributes equitably the
heads that are unaccounted for to all entities for which the counts are
missing. It respects the bounds provided as input.

\begin{algorithm}[H]
\caption{\textsc{FillGaps} integer program to fill missing gaps in state
and county totals and state counts by farm size.} 
\label{alg:fillgap}
\KwIn{No. of unknown quantities~$m$, their sum total~$T$, and bounds
$\big((L_1,U_1)$, $(L_2,U_2)$, $\ldots$, $(L_m,U_m)\big)$,
where $L_i \leq U_i$, $1 \leq i \leq m$.}
\KwOut{Assignment of values to the $m$ unknown quantities.
(The constraints to be satisfied by the unknown quantities
are provided below.)}
\BlankLine
\textbf{Variables}\\
$x_i$, $i=1,\ldots,m$ \hfill \tcp{Variables for unknown quantities}
$\lambda_0 \geq 0$ \hfill \tcp{Variable for equitable distribution}
\BlankLine
\textbf{Constraints}\\
$L_i\le x_i \le U_i$, $1\le i\le m$ \hfill 
\tcp{Bounds on unknown quantities based on input data}
$\sum_i x_i = T$ \hfill \tcp{Sum of unknown quantities should be T}
$x_i - L_i \leq \lambda_0$ \hfill \tcp{Bound the difference between the 
assigned quantities and corresponding lower bounds}
\BlankLine
\textbf{Set Objective:} Minimize $\lambda_0$\\
\BlankLine
\Return $(x_1,x_2,\ldots,x_m)$
\end{algorithm}

Now, we describe the process used to fill missing data for each count type
in the order in which they are processed.

\begin{enumerate}
\item \textbf{State-total head counts.}
The total head count here corresponds to the country head count that is
available for every livestock subtype. There are potentially four sources
that can be used to derive bounds. If farm counts per farm-size category are
given at the state level, an initial set of lower and upper bounds can be
derived as follows: $\wimin F_{i\gamma}\le H_{i\gamma}\le \wimax
F_{i\gamma}$. The lower bounds are refined by using the available head
counts for each farm size category. Similar refinement can be done from
counts per farm size at the county level. The sum of the lower bounds
across farm categories provides a lower bound for the state total. Finally,
if county totals are provided for some counties of the state, their sum
provides another lower bound. We set the final lower bound to be the
maximum of bounds obtained as above. This data is fed to \textsc{FillGaps}
to obtain estimates for the missing counts.

\item \textbf{State-by-farm-size head counts.}
The total head count here corresponds to state total which is either
available or estimated as above for every livestock subtype. We use the
same approach as above by first deriving bounds based on the number of farms
in each category and then refining the lower bound using
county-by-farm-size head counts. This data is fed to \textsc{FillGaps} to
obtain estimates for the missing counts.

\item \textbf{County-total head counts.}
The total head count here corresponds to the total head count in the state
to which the county belongs, which is either available or estimated as
above for every livestock subtype. If farm counts are provided for each
farm category, then we use it to derive the initial bounds. This is
further refined if county-by-farm-size head counts are provided. This data
is fed to \textsc{FillGaps} to obtain estimates for the missing counts.

\item \textbf{County-by-farm-size head counts using IPF.} To fill gaps in
county-by-farm-size counts, we follow the methodolgy of
Burdett~et~al.~\cite{burdett2015simulating}. They apply
IPF~\cite{fienberg1970iterative,deming1940least} for the case of hogs to
estimate these counts. Here, we give an overview of the method and refer
to Burdett~et~al.~\cite{burdett2015simulating} for details. The processing is done per state and
subtype. In the IPF process, the objective is to impute missing values in a
given matrix given row and column totals. In this case, the data matrix
consists of county-by-farm-size counts with counties as rows and farm size
categories as columns. Note that, at this stage, both county totals and
state-by-farm-size counts are available either from data or by estimation.
Unknown values in the matrix are seeded with the product of the average
size of the corresponding category and the number of farms in that
category.
\end{enumerate}

\subsection{Generation of farms}
\label{sec:gen_farms}

\paragraph{Objective.} We are given farm categories~$i=1,\ldots,\ell$.
Let~$\Fi$ and~$\Hi$ denote the number of farms and livestock heads in
category~$i$. For~$k=1,\ldots,\ell$, let~$\Fgk$ and~$\Hgk$ denote the
number of farms and heads corresponding to category~$k$ with respect to
subtype~$\gamma$ (see Table~\ref{tab:ftoc_input}). The objective is to find
an assignment of farms whose farm counts and composition respects these
counts. Note that the categories are pairwise disjoint and cover the entire
range. In addition to head counts of subtypes, we are also given counts of
the livestock type as well by farm size. Since the IPF procedure does not
account for these counts, there could be differences between this data and
the counts obtained by aggregating farm subtype counts in our assignment.
Our optimization objective is a linear combination of many
parameters as discussed below.
\begin{description}
\item{(a)} To choose an assignment with minimum discrepancy with respect to known
counts, we introduce a parameter~$\lambda_1$.

\item{(b)} The parameter $\lambda_2$ in the 
minimization objective represents the largest
number of subtypes in a farm.

\item{(c)} The minimization objective includes $\ell$ parameters, denoted by
$\lambda_{3i}$, $1 \leq i \leq \ell$.
The purpose of parameter $\lambda_{3i}$ is to ensure that the 
population of all the subtypes in any farm of category $i$
is close to the average value for that farm category.
The purpose of minimizing these parameters is to obtain
an equitable distribution of the population
across all farm categories.

\item{(d)} The parameter $\lambda_4$ in the minimization
objective is used to ensure that
the sum of the head counts of subtype $\gamma$ over all the farms assigned category $k$ is close to the given head count 
$H_{\gamma k}$.
\end{description}
The optimization objective combines the above parameters into 
a linear function using appropriate scaling constants.
These scaling constants ensure that parameters with larger 
values have larger penalties.
As a consequence, the solver will aggressively minimize
parameters with larger values compared to ones with smaller values.

\begin{algorithm}[H]
\small
\caption{\textsc{GenFarms}. Integer linear program to generate farms consistent
with input counts of farms and head counts.} 
\label{alg:gen_farms}
\KwIn{County-by-farm-size head and farm counts as in
Table~\ref{tab:ftoc_input}.}
\KwOut{Farms with head counts of each subtype.}
\BlankLine
\textbf{Variables}\\
For each farm~$f$ in category~$i$, $\hifg$ corresponds to the
population of subtype~$\gamma$ in that farm\;
For each farm~$f$ in category~$i$, $\xifgk$ indicates whether~$(f,i)$
belongs to category~$k$ w.r.t. subtype~$\gamma$\;
For each farm~$f$ in category~$i$, $\yifgk$ indicates whether~$\hifg=0$\;
For each farm~$f$ in category~$i$, $\zifgk=\hifg$ if $\hifg$ belongs
to the category~$k$ with respect to subtype~$\gamma$. Otherwise, it
is~$0$.\\
Variables for minimization objectives: $\lambda_1$ (alignment with known
total population within each farm size category), $\lambda_2$ (minimize
number of subtypes per farm),
$\lambda_{3i}$,~ $i\in[1,\ell]$ (equitable distribution of population in
each farm size category), and (alignment with subtype population within
each farm size category) $\lambda_4$.
\BlankLine
\textbf{Constraints}\\
Let~$M$ be a suitably large constant\;
\tcp{Population and farm size constraints}
$\wimin\le\sum_\gamma\hifg\le\wimax$ \hfill \tcp{Category farm size constraint}
\BlankLine
\tcp{Subtype farm category constraints: Farm counts.}
$i\in[1,\ell],f\in[1,F_i],\gamma\in\Gamma,k\in[1,\ell]$,~ 
$\hifg\ge\wkmin-(1-\xifgk)\cdot M$ \hfill\tcp{Lower bound}
$i\in[1,\ell],f\in[1,F_i],\gamma\in\Gamma,k\in[1,\ell]$,~ 
$\hifg\le\wkmax+(1-\xifgk)\cdot M$ \hfill\tcp{Upper bound}
$i\in[1,\ell],f\in[1,F_i],\gamma\in\Gamma,k\in[1,\ell]$,~ 
$\hifg\ge1-\yifgk$ \hfill\tcp{Lower bound when subtype count is 0}
$i\in[1,\ell],f\in[1,F_i],\gamma\in\Gamma,k\in[1,\ell]$,~ 
$\hifg\le(1-\yifgk)\cdot M$ \hfill\tcp{Upper bound when subtype count is 0}

$i\in[1,\ell],f\in[1,F_i],\gamma\in\Gamma$,~ 
$\sum_{k}\xifgk+\yifgk = 1$\hfill \tcp{Farm in exactly one category}
$\gamma\in\Gamma,k=[1,\ell]$,~ 
$\sum_{f,i}\xifgk = \Fgk$\hfill \tcp{Count equals total farms in that
category}
\BlankLine
\tcp{Subtype farm category constraints: Population counts}
$i=[1,\ell],~f=[1,F_i],~\gamma\in\Gamma,~k=[1,\ell]$,~ 
$\zifgk\le\hifg$ \hfill\tcp{Upper bound}
$i=[1,\ell],~f=[1,F_i],~\gamma\in\Gamma,~k=[1,\ell]$,~ 
$\zifgk\le\xifgk\cdot M$\hfill \tcp{$\xifgk=0\Rightarrow\zifgk=0$}
$i=[1,\ell],~f=[1,F_i],~\gamma\in\Gamma,~k=[1,\ell]$,~ 
$\zifgk\ge\hifg-(1-\xifgk)\cdot M$ \hfill \tcp{$\xifgk=1\Rightarrow\zifgk=\hifg$}
$\gamma\in\Gamma,~k=[1,\ell]$,~
$\Big|\sum_{f,i}\zifgk-\Hgk\Big|\le \lambda_4$\hfill \tcp{Count must be close to $\Hgk$}
\BlankLine
\tcp{The assignment should be such that it is as close a match to the total
population distribution in each category,~$\Hi$.}
$i=[1,\ell],~\Big|\sum_{f,\gamma}\hifg-\Hi\Big|\le\lambda_1.$
\BlankLine
\tcp{Subtype distribution: Minimize number of subtypes per farm}
$i=[1,\ell],~f=[1,F_i]$,~$\sum_{\gamma,k}\xifgk\le\lambda_2$\\
\BlankLine
\tcp{Equitable distribution in each category.}
$i=[1,\ell],~a_i=\sum_{f,\gamma}\hifg/F_i$,\hfill\tcp{Average population in
each farm category}
$i=[1,\ell],~f=[1,F_i],~\Big|\sum_{\gamma}\hifg-a_i\Big|\le\lambda_{3i}$\\
\BlankLine
{\bf Set objective.}
Minimize~$\lambda_1+(H+1)\cdot\lambda_2+(|\Gamma|+1)(H+1)\cdot
\sum_i\lambda_{3i} + (|\Gamma|+1)(H+1)^2\cdot\lambda_4$.
\BlankLine
\Return $(\hifg\mid i=[1,\ell],f=[1,F_i],\gamma\in\Gamma)$
\end{algorithm}

\paragraph{Implementation notes.} The algorithm was run for each
county--livestock instance in parallel on a HPC cluster. For faster
convergence to a solution, we set the MIP gap (which refers to the percentage
difference between the current best feasible solution and the best known
bound on the optimal objective value) to $0.1\%$ of the total head count
for each instance.

\subsection{Farms to cells}\label{sse:farms_to_cells}
\paragraph{Objective.} We are given~$N_f$ farms with number of heads in
each farm~$f_i$ denoted by~$P_i$, $i=1,\ldots, N_f$, and~$N_c$ cells with
number of heads in each cell~$C_j$ denoted by~$Q_j$, $j=1,\ldots,N_c$. The
objective is to assign to each farm~$f_i$ a cell~$C_j$ such that~$\max_j
\big|\sum_{i,\delta_i=j} P_i- Q_j\big|$ is minimized, where ~$\delta_i=j$
if and only if~$f_i$ is assigned cell~$C_j$.

\begin{algorithm}[H]
\caption{\textsc{FarmsToCells}. Integer linear program to generate farms
consistent with input counts of farms and head counts.}
\label{alg:farms_to_cells}
\KwIn{Farms $f_i$, $i=[1,N_f]$ with total head count~$P_i$, cells $C_j$,
$j=[1,N_c]$ with head count~$Q_j$.}
\KwOut{Assignment of each farm to a cell.}
\BlankLine
\textbf{Variables}\\
For~$1\le i\le N_f$ and~$1\le j\le N_c$,~$x_{ij}$ indicates whether
farm~$f_i$ is assigned to cell~$C_j$:~$x_{ij}=1$ if~$f_i$ is assigned
cell~$C_j$; otherwise, $x_{ij} = 0$\;
For~$1\le i\le N_f$ and~$1\le j\le N_c$,~$h_{ij}=P_i$ if~$x_{ij}=1$,
else~$0$\;
Let~$\lambda_4$ be a positive integer variable that is equal to
$\max_j \big|\sum_{i,\delta_i=j} P_i- Q_j\big|$\;
\BlankLine
\textbf{Constraints}\\
\tcp{Assign farm to a cell}
$1\le i\le N_f, 1\le j\le N_c$,~ $h_{ij} = x_{ij}\cdot
P_i$\hfill\tcp{Contribution of a farm to cell population}
$1\le i\le N_f$,~ $\sum_{j}x_{ij}=1$\hfill\tcp{Each farm belongs to exactly
one cell}
$1\le j\le N_c$,~
$\big|\sum_ih_{ij}-Q_j\big|\le\lambda_5$\hfill\tcp{Farm assignment should
align with cell capacities}
\BlankLine
{\bf Set objective.} Minimize~$\lambda_5$.
\BlankLine
\Return $(\hifg\mid i=[1,\ell],f=[1,F_i],\gamma\in\Gamma)$
\end{algorithm}

\paragraph{Implementation notes.} The algorithm was run for each
county--livestock instance in parallel on a HPC cluster. For faster
convergence to a solution, we set the MIP gap to $0.01\%$ of the total head
count for each instance.

\subsection{Additional results}
\begin{figure}[htb]
\centering
\begin{subfigure}[b]{.49\textwidth}
\includegraphics[width=\textwidth]{figs/livestock_cafo_100_10000_200.pdf}
\end{subfigure}
\begin{subfigure}[b]{.49\textwidth}
\includegraphics[width=\textwidth]{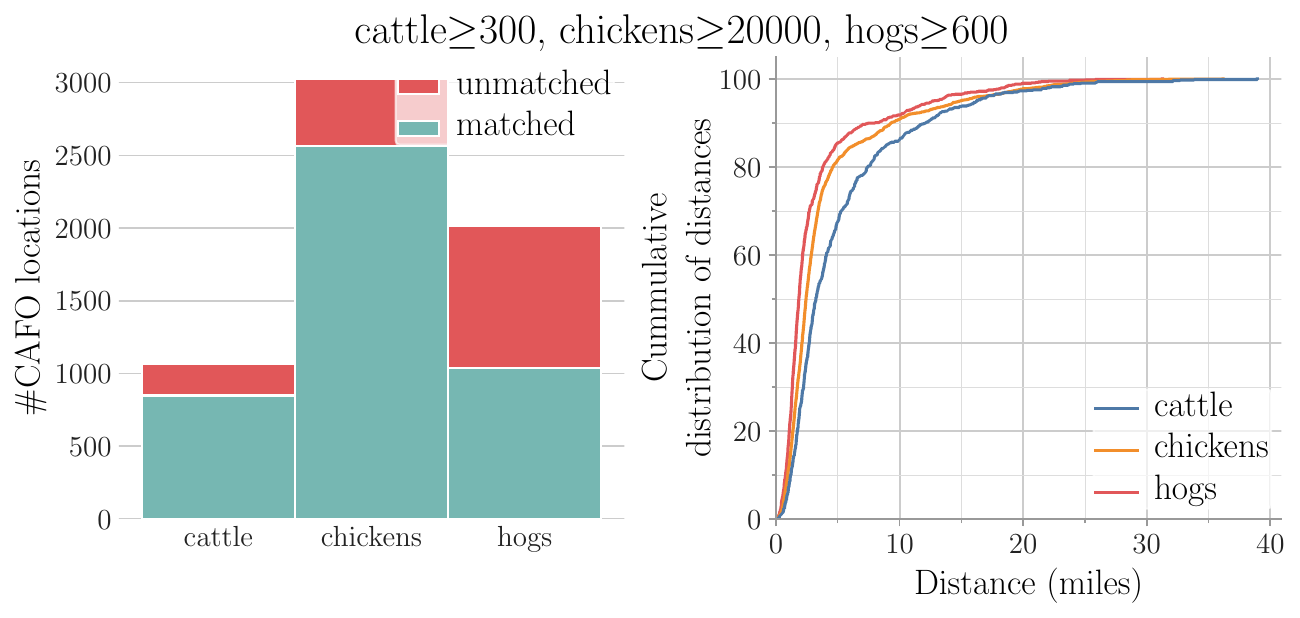}
\end{subfigure}
\caption{Analysis of mapping CAFO locations by livestock type to farms from
the digital similar. The two plots correspond to two sets of thresholds for
choosing the farms to compare with. The second set corresponds to larger
farms compared to the first. In each plot, the first subplot shows how
many CAFO locations were matched. The second subplot provides the
cumulative distribution of the distances~(in miles) between matched pairs
of CAFO locations and farms.}
\label{fig:additional_cafo}
\end{figure}

\section{Wild Birds}
Our processing pipeline to extract abundance data involves the following steps:
\begin{itemize}
\item Coordinate System Conversion: We transform the data from its original
World Eckert IV equal-area projection (ESRI:54012) to a standard geographic
coordinate system (EPSG:4326) to ensure compatibility with other geospatial
datasets in our study.
\item Spatial Sampling: To balance spatial resolution with computational
efficiency, we sampled the data at regular intervals consistent with the
\glw{} grid cell dimensions. This sampling strategy preserves the overall
spatial patterns while reducing the dataset to a manageable size.
\item Temporal Resolution: We maintained the weekly temporal resolution
provided by eBird, allowing for detailed analysis of seasonal migration
patterns.
\end{itemize}


\begin{figure}
    \centering
    \includegraphics[width=0.7\textwidth]{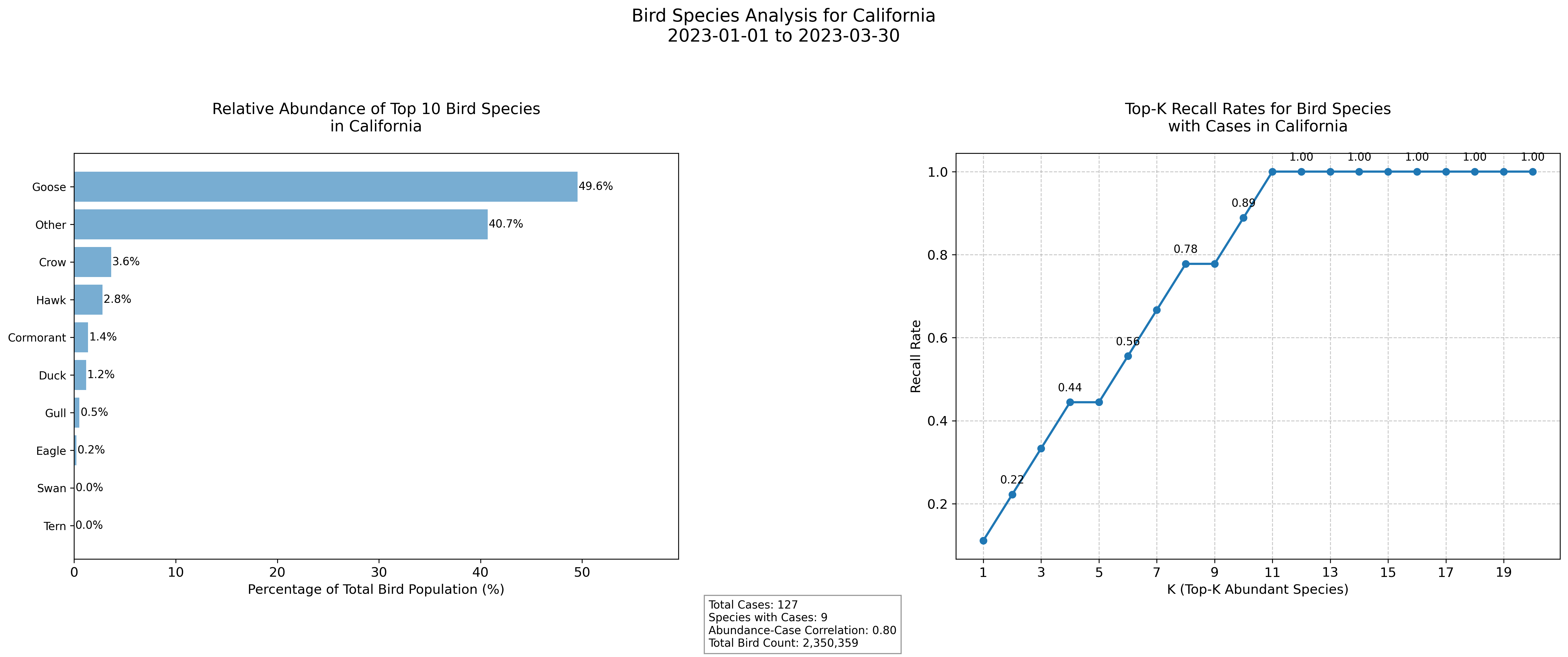}
    \includegraphics[width=0.7\textwidth]{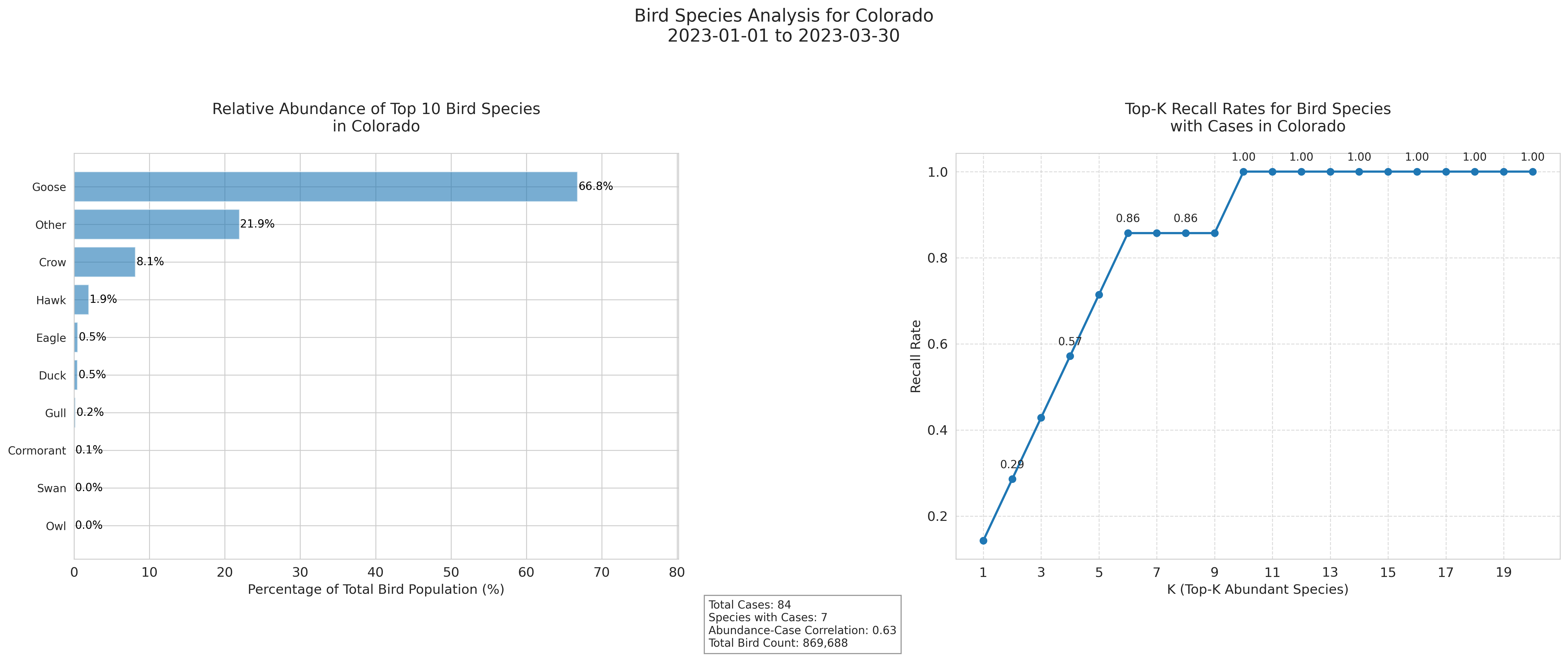}
    \includegraphics[width=0.7\textwidth]{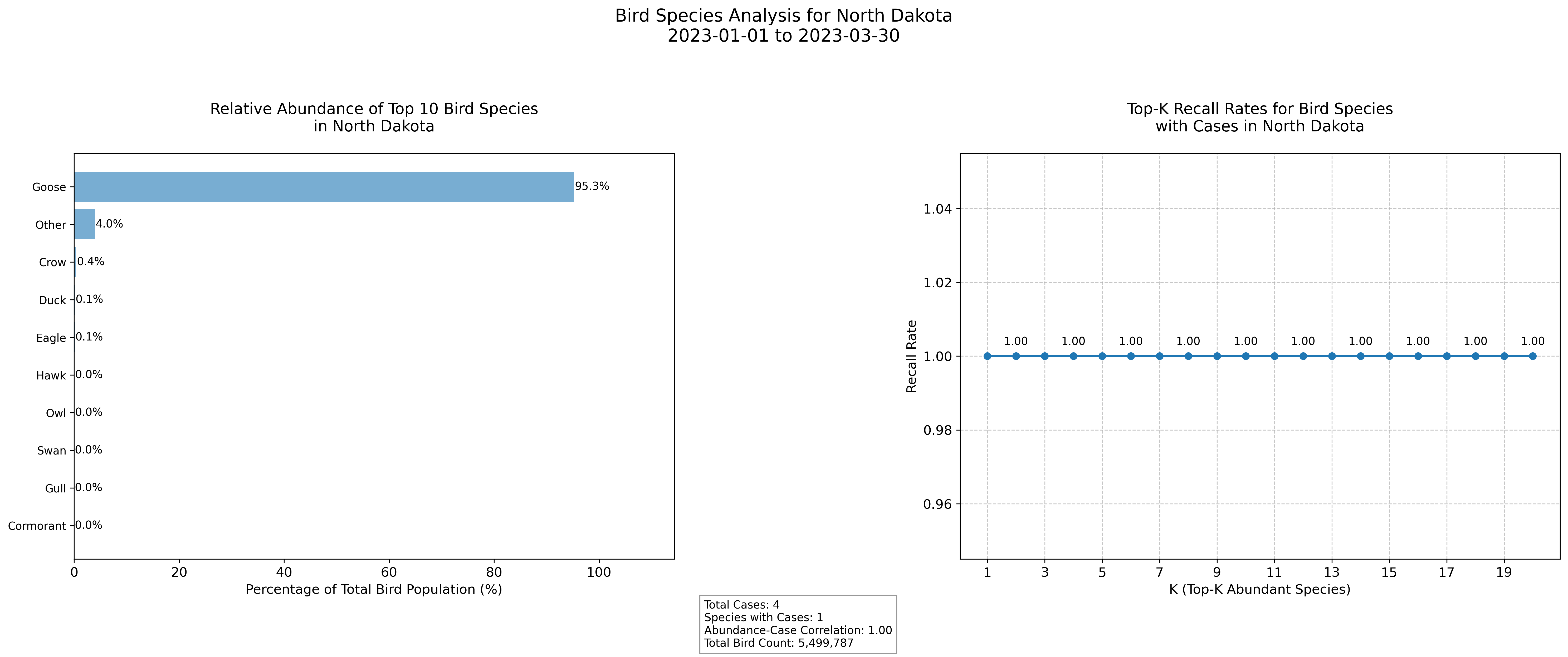}
    \includegraphics[width=0.7\textwidth]{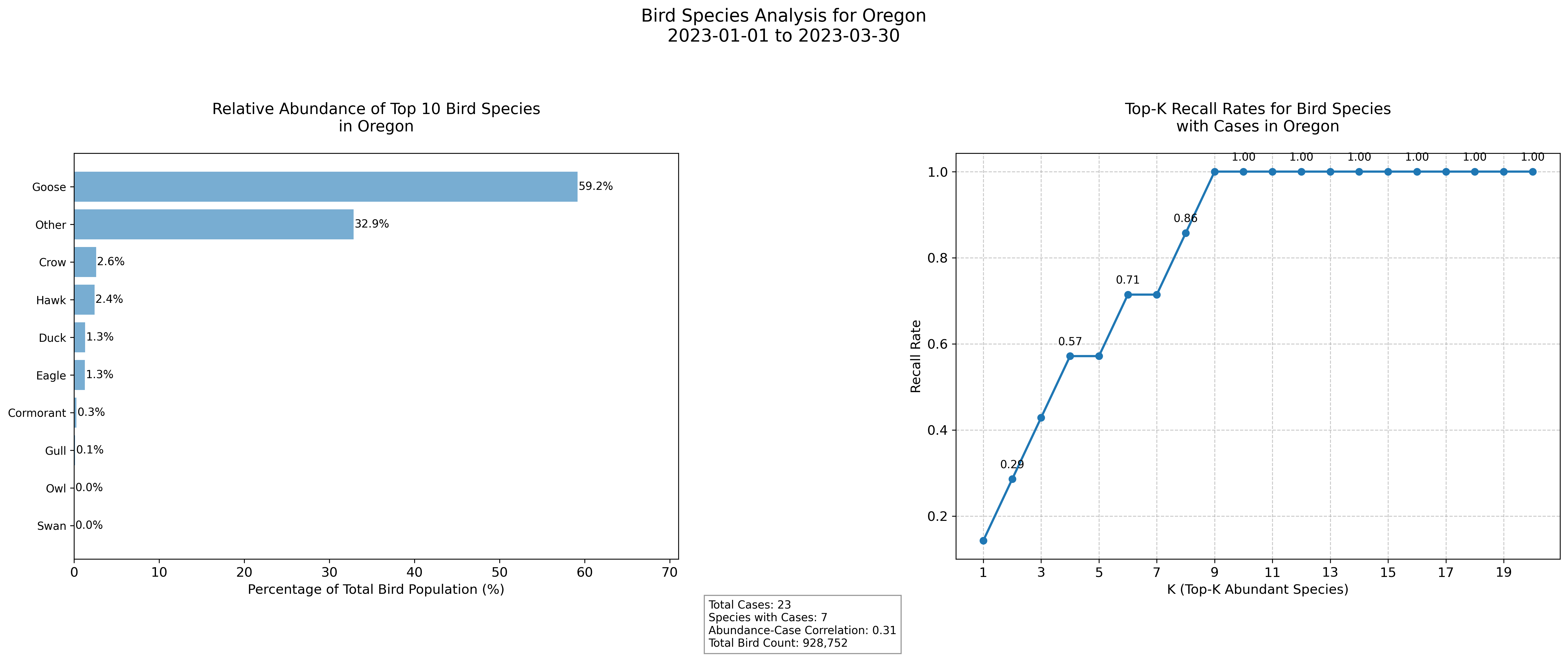}
    \caption{Bird Abundance and H5N1 incidence data from January to March 2023.}
    \label{fig:recall_fig}
\end{figure}


\begin{figure}
    \centering
    \includegraphics[width=0.48\linewidth]{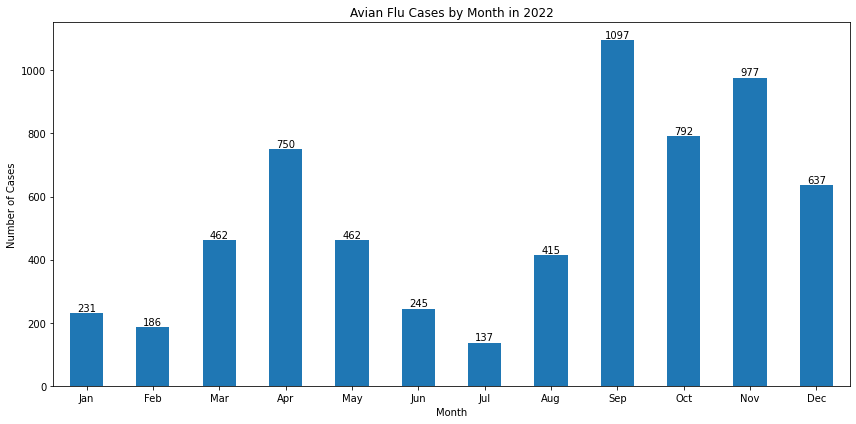}
    \includegraphics[width=0.48\linewidth]{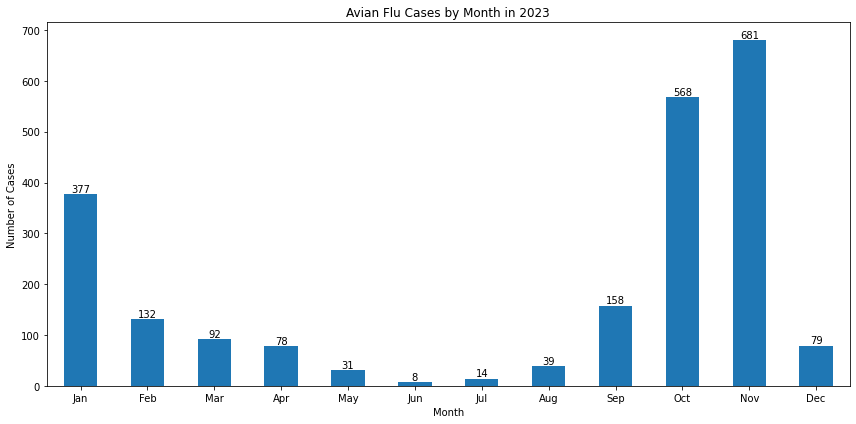}
    \caption{H5N1 incidence in wild birds across different months in 2022 and 2023. Case abundance is higher in fall and winter months, which can be attributed to breeding and migration patterns and viral transmissibility in colder months.}
    \label{fig:h5n1_incidence_by_month}
\end{figure}

\begin{figure}
    \centering
    \includegraphics[width=0.80\linewidth]{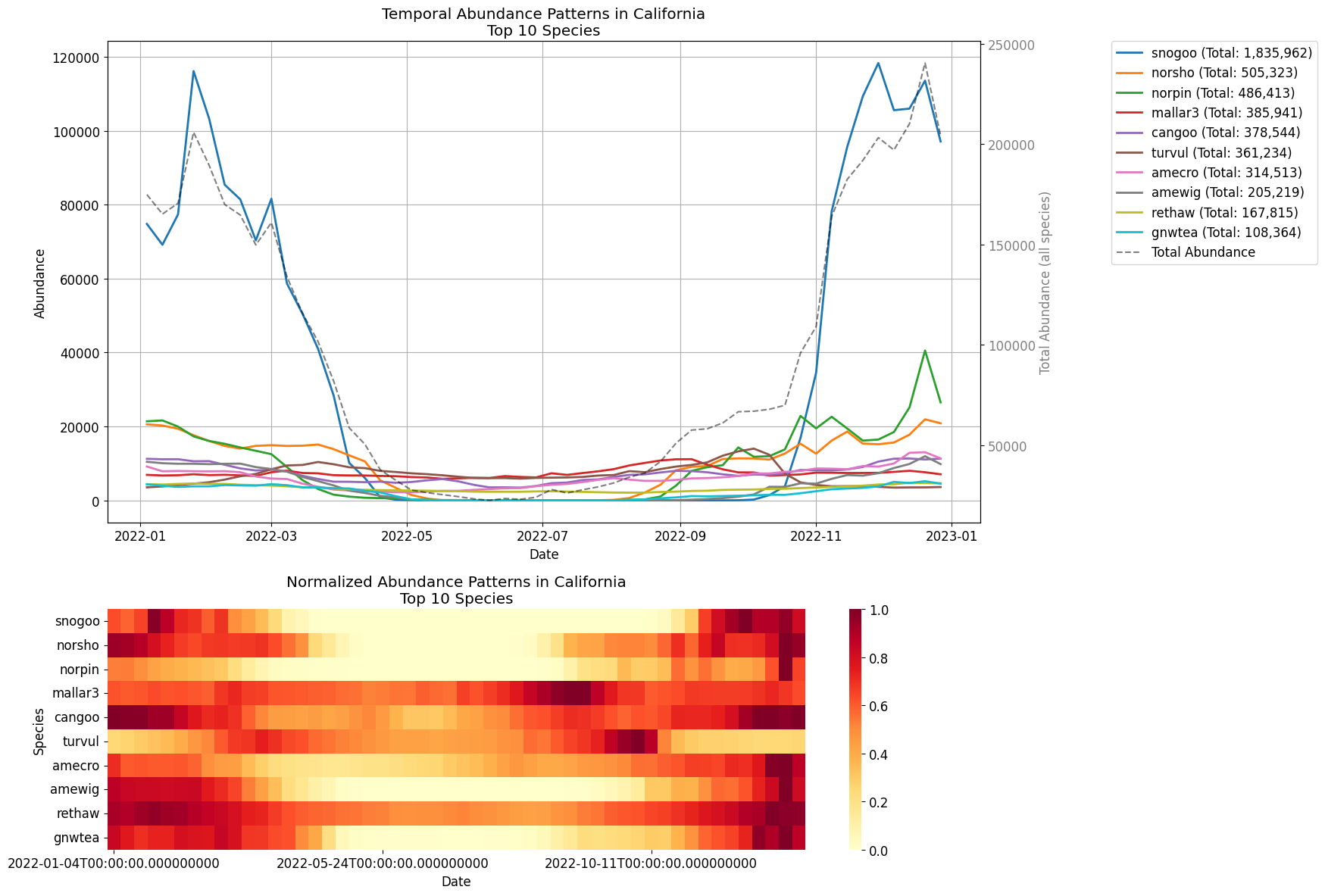}
    \includegraphics[width=0.80\linewidth]{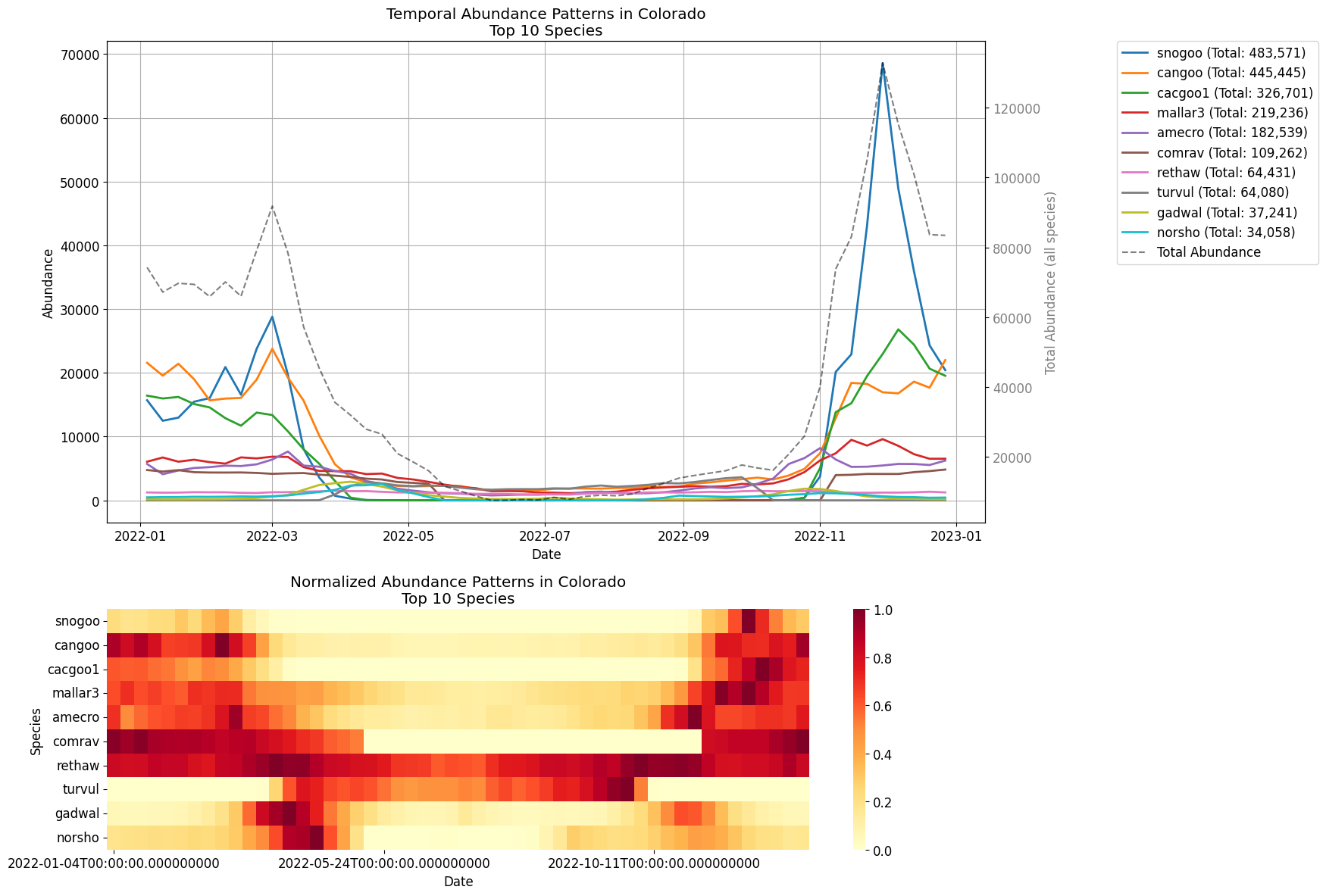}
    \caption{Temporal Abundance Pattern of Bird Species across states. We observe heterogeneity of species abundance and demographics across different states, throughout the year. Abundance varies across seasons due to migration and breeding in different geographies.}
    \label{fig:abundance1}
\end{figure}

\begin{figure}
    \centering
    \includegraphics[width=0.80\linewidth]{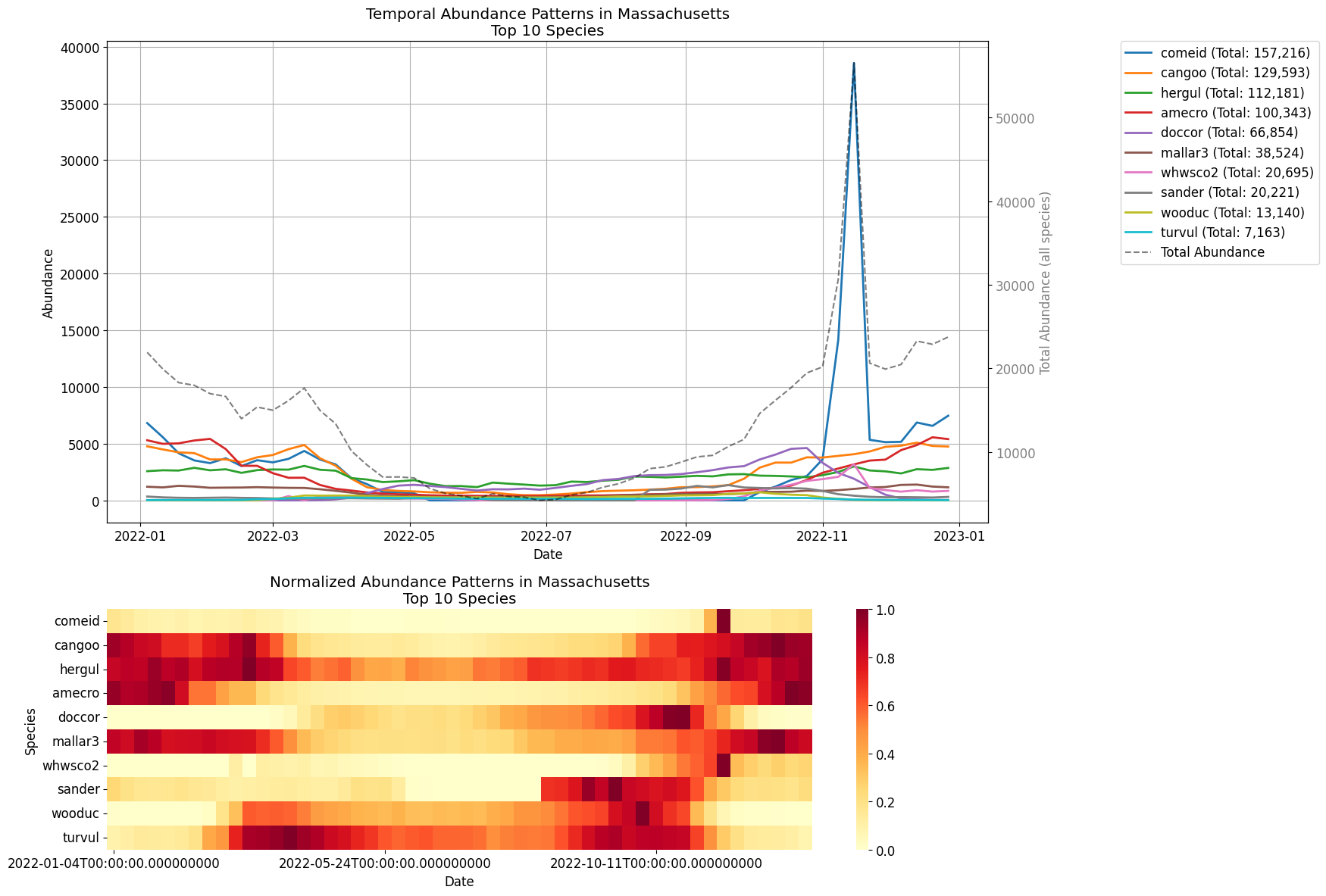}
    \includegraphics[width=0.80\linewidth]{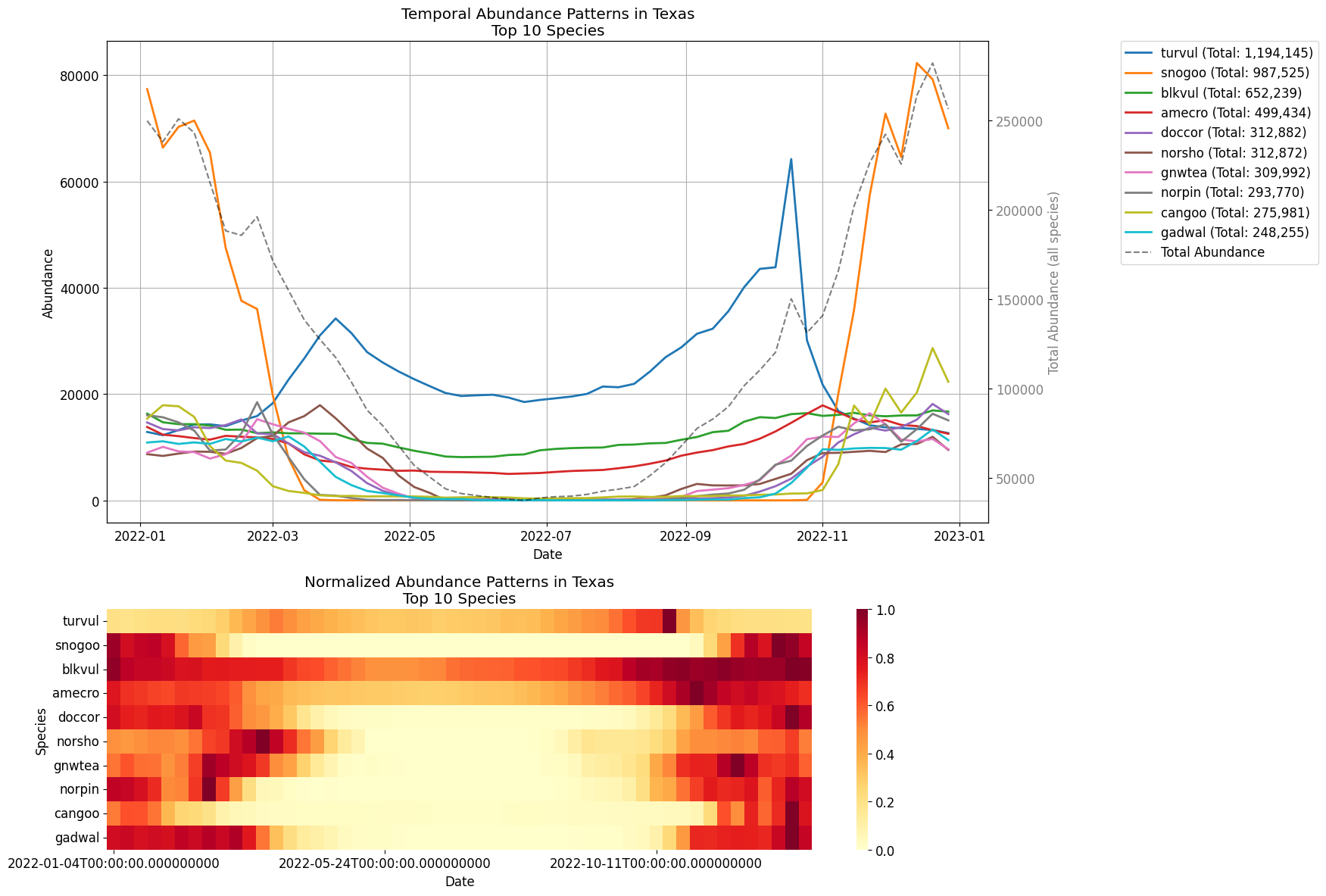}
    \caption{Temporal Abundance Pattern of Bird Species across states. We observe heterogeneity of species abundance and demographics across different states, throughout the year. Abundance varies across seasons due to migration and breeding in different geographies.}
    \label{fig:abundance2}
\end{figure}

\begin{figure}
    \centering
    \includegraphics[width=0.80\linewidth]{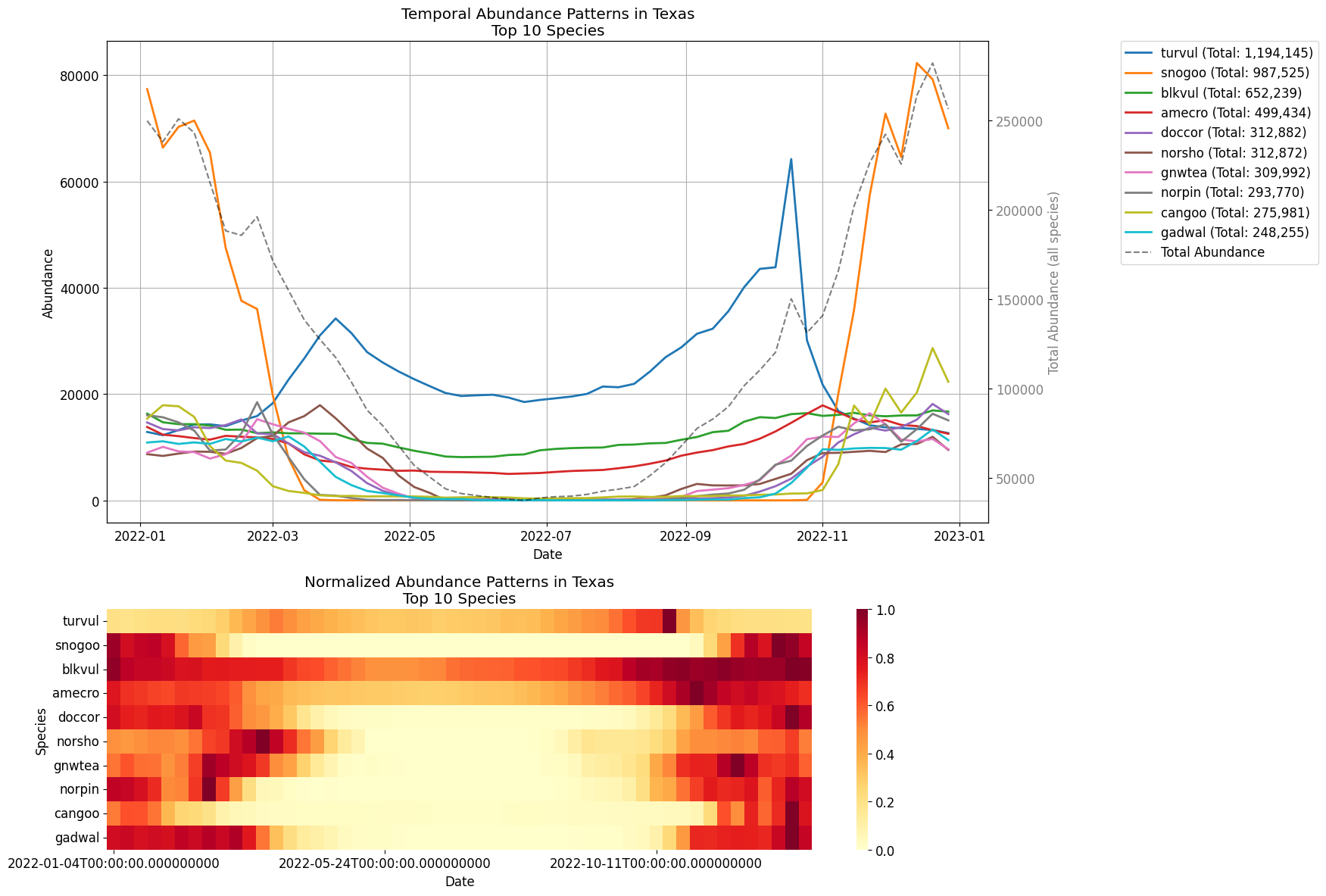}
    \caption{Temporal Abundance Pattern of Bird Species across states. We observe heterogeneity of species abundance and demographics across different states, throughout the year. Abundance varies across seasons due to migration and breeding in different geographies.}
    \label{fig:abundance3}
\end{figure}

\section{Additional Information on Processing Centers}
\label{ssec:dairy}

\begin{table}[htb]
\centering
\begin{tabular}{c|c|p{7cm}}
\textbf{Risk Level} & \textbf{Product Codes}     & \textbf{Description}                                                                                  \\ \hline
High Risk           & M1, M2, M3, M6, M7, M8, M9 & Raw milk, cream, and concentrated milk products            \\ \hline
High Risk           & C3-C47                     & Cheese types use unpasteurized milk \\ \hline
High Risk           & B1-B9                      & Butter products, made from unpasteurized cream                                 \\ \hline
Medium Risk         & M10, M11, M12, M13, M14    & Processed milk products involving some heat treatment                 \\ \hline
Low Risk            & D1-D18                     & Dry milk products, undergo heat treatment during processing                           \\ \hline
Low Risk            & W1-W25                     & Whey products, derived from pasteurized milk during cheese-making                   \\ \hline
Low Risk            & F1-F15                     & Frozen dessert products, made with pasteurized ingredients                                  \\ \hline
Low Risk            & S4-S46                     & Specialty products which involve processing that would eliminate pathogens                   \\ \hline
\end{tabular}
\caption{Risk of engaging with unpasteurized milk for different plant codes.}
\label{tab:plant_codes}
\end{table}

The USDA~\cite{usdadairy}
maintains a comprehensive list of approved dairy plants, including
pasteurization facilities, which are inspected at least twice yearly. These
inspections cover over 100 items, including milk supply, plant facilities,
equipment condition, sanitary practices, and processing procedures. While
most commercial dairy operations use pasteurized milk due to food safety
regulations, some facilities may handle unpasteurized milk for specific
products or processes.

The USDA assigns dairy plant approvals using two categories:
\begin{itemize}
    \item Section I: Plants that produce products manufactured from dairy
    ingredients meeting USDA requirements or originating from USDA-approved
    plants. These are generally considered lower risk in terms of handling
    unpasteurized milk. (designated B, C, D, F, M, S and W codes)
    \item Section II: Plants that may have products produced from dairy
    ingredients that did not originate from USDA-approved plants
    (unapproved source plants). These plants potentially pose a higher risk
    of handling unpasteurized milk. (designated P codes)
\end{itemize}

We categorize product codes based on their potential risk of interaction with unpasteurized milk in Table~\ref{tab:plant_codes}.

\section{Additional results for risk estimation}

\begin{figure}
    \centering
    \begin{subfigure}[b]{\linewidth}
        \includegraphics[width=.24\textwidth]{figs/colocation_time_agg_risk_map_milk.pdf}
        \includegraphics[width=.24\textwidth]{figs/colocation_time_agg_risk_map_turkeys.pdf}
        \includegraphics[width=.24\textwidth]{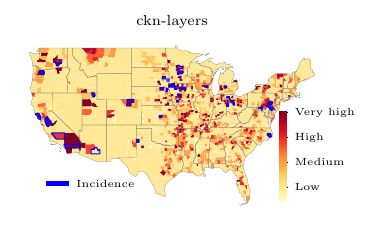}
        \includegraphics[width=.24\textwidth]{figs/colocation_time_agg_risk_map_cattle.pdf}
        \caption{Elevated Risk Level}
    \end{subfigure}
    \begin{subfigure}[b]{\linewidth}
        \includegraphics[width=.24\textwidth]{figs/colocation_time_peak_risk_map_milk.pdf}
        \includegraphics[width=.24\textwidth]{figs/colocation_time_peak_risk_map_turkeys.pdf}
        \includegraphics[width=.24\textwidth]{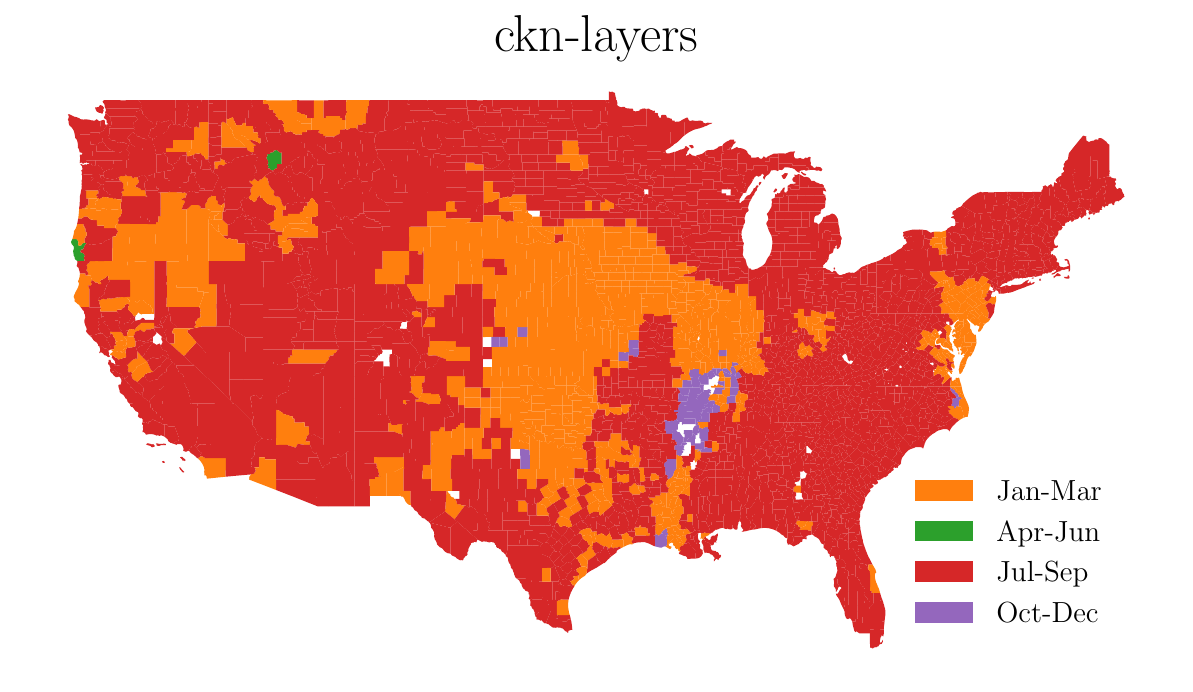}
        \includegraphics[width=.24\textwidth]{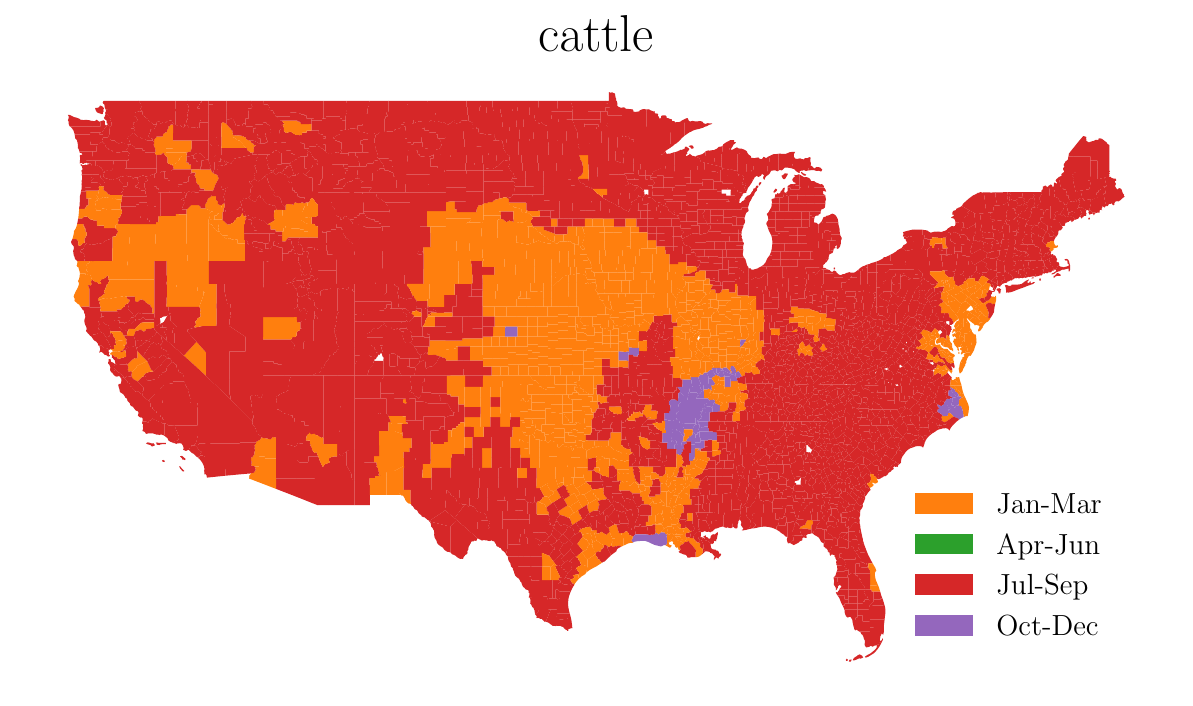}
        \caption{Period of peak risk}
    \end{subfigure}
    \begin{subfigure}[b]{\linewidth}
        \includegraphics[width=.32\textwidth]{figs/colocation_recall_milk.pdf}
        \includegraphics[width=.32\textwidth]{figs/colocation_recall_turkeys.pdf}
        \includegraphics[width=.32\textwidth]{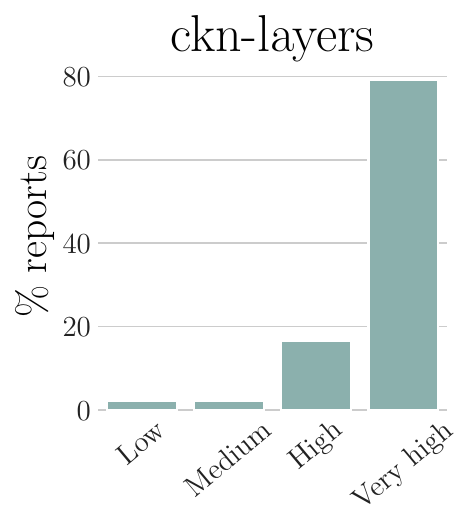}
        \caption{Recall performance.}
    \end{subfigure}
    \caption{Summarizing colocation risk. (a)~Aggregated risk levels across
    counties using stable sorting: Counties are ordered by the highest severity of
    risk among all quarters and then among all counties with the same
    highest severity, they are ordered by the number of quarters in which
    this severity occurs. We continue this across all severities. (b)~Peak
    time of risk: The quarter with the highest risk for each county is
    plotted. (c) Recall performance when colocation risk is compared with
    ground truth.}
    \label{fig:agg_risk_maps}
\end{figure}

\begin{figure}
\centering
\begin{subfigure}[b]{\textwidth}
    \centering
    \includegraphics[width=\textwidth]{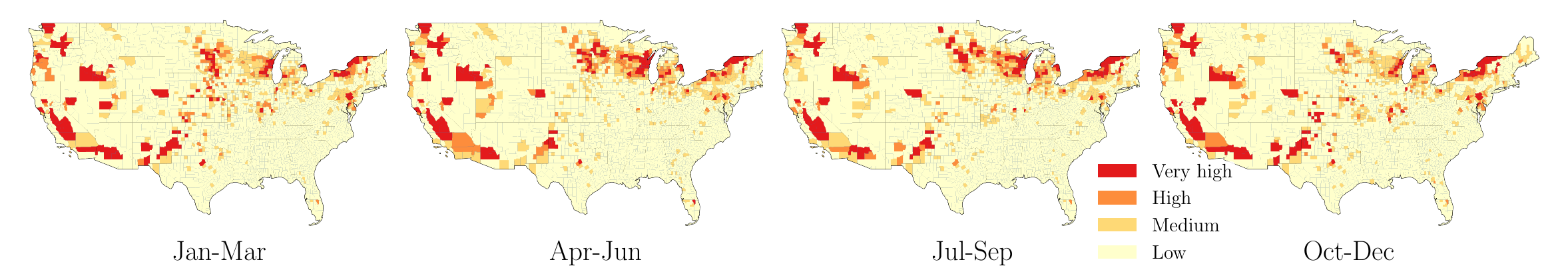}
    \caption{Dairy cattle}
    \label{figs:riskmaps:dairy}
\end{subfigure}
\begin{subfigure}[b]{\textwidth}
    \centering
    \includegraphics[width=\textwidth]{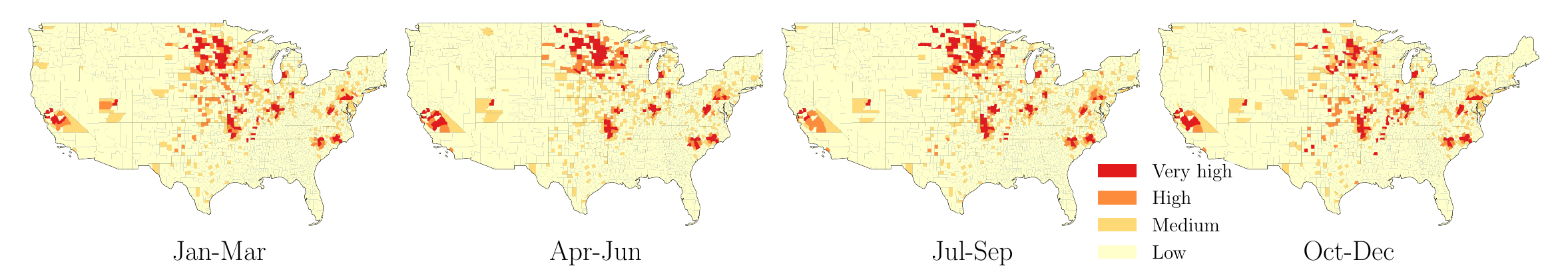}
    \caption{Turkeys}
    \label{figs:riskmaps:turkeys}
\end{subfigure}
\begin{subfigure}[b]{\textwidth}
    \centering
    \includegraphics[width=\textwidth]{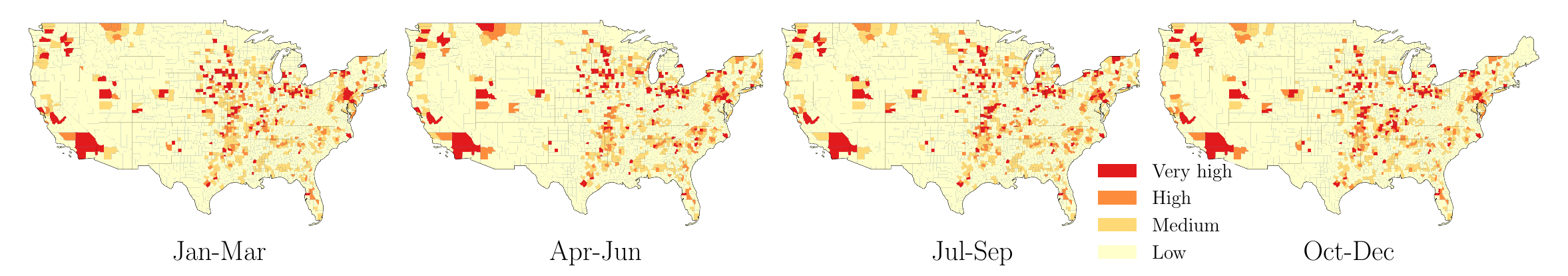}
    \caption{Chicken layers}
    \label{sup:fig:riskmaps:layers}
\end{subfigure}
\begin{subfigure}[b]{\textwidth}
    \centering
    \includegraphics[width=\textwidth]{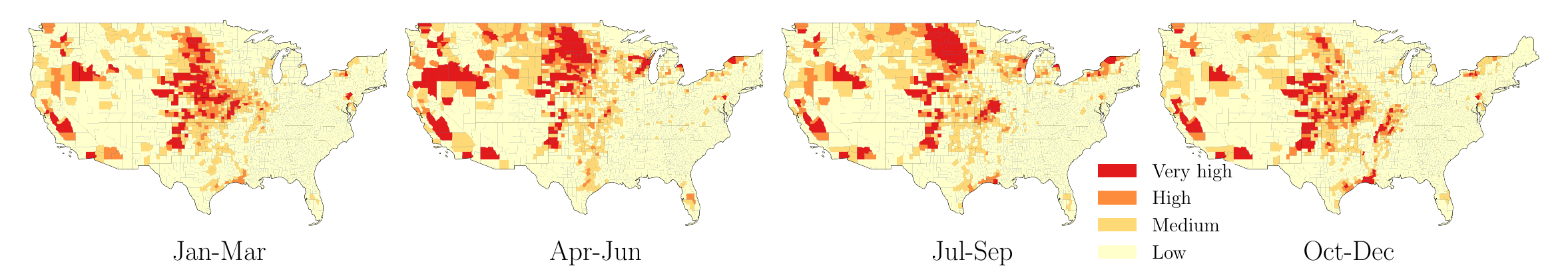}
    \caption{Cattle}
    \label{sup:fig:riskmaps:cattle}
\end{subfigure}
\caption{Collocation risk for the four quarters of the year for several
affected livestock subtypes.}
\label{sup:fig:cattle}
\end{figure}

\begin{figure}
    \centering
    \includegraphics[width=0.9\linewidth]{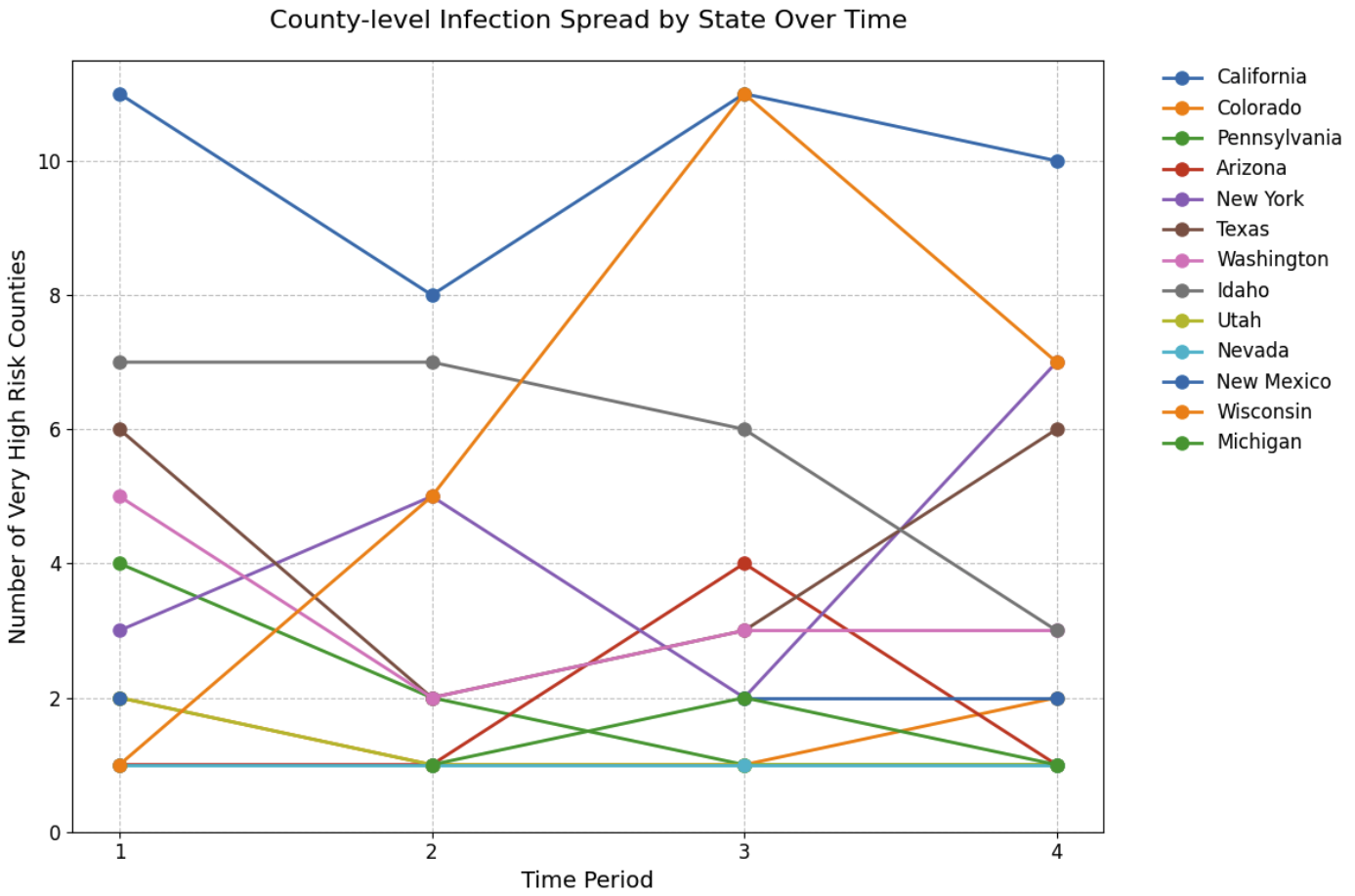}
    \caption{Time series showing the number of very high risk counties
    (95th percentile) in each state across four periods in 2024:
    January-February, March-May, June-August, and September-December. Only
    states that had at least one very high risk county in three or more
    periods are displayed. The seasonal variation across states can be explained due to the migration patterns of wild birds. While some states, like Nevada and Utah, had fewer very high risk counties, their persistent presence across multiple periods suggests sustained elevated risk in specific regions throughout the year.}
    \label{fig:bump-chart-dairy-cattle}
\end{figure}

\begin{table}[htbp]
    \centering
    \caption{We visualize risk persistence and observed outbreaks in counties that consistently showed very high risk (95th percentile) across time periods in 2024. In this table, Avg. Position indicates the mean ranking of risk level across appearances, with lower numbers indicating higher risk (e.g., position 1 means highest risk in that period). Time periods are: Jan-Feb, Mar-May, Jun-Aug, and Sep-Dec 2024. Counties appearing in all time periods demonstrate persistent risk factors throughout the year, regardless of seasonal variations. We validate our results against real-world outbreak instances from WHO and CDC reportings. }
    \small
    \begin{tabular}{l|l|l|l|l|l}
        \hline
        \textbf{County} & \textbf{State} & \textbf{Appearances} & \textbf{Avg. Position} & \textbf{Variance} & 
        \textbf{Known Outbreak} \\
        \hline
        Merced & CA & 4 & 1.5 & 0.2 & Yes\\
        Tulane & CA & 4 & 3.0 & 1.5 & Yes \\
        Weld & CO & 4 & 3.5 & 10.2 & Yes \\
        San Joaquin & CA & 4 & 4.2 & 2.7 & Yes \\
        Kern & CA & 4 & 5.8 & 0.7 & Yes \\
        Fresno & CA & 4 & 7.5 & 6.8 &Yes \\
        Maricopa & AZ & 4 & 8.0 & 3.5 & Yes \\
        Kings & CA & 4 & 9.2 & 2.7 & Yes \\
        Stanislaus & CA & 4 & 9.8 & 5.2 & ToDo \\
        Lancaster & PA & 4 & 13.5 & 58.8 & Yes\\
        San Bernardino & CA & 4 & 14.8 & 6.2 & Yes\\
        Box Elder & UT & 4 & 15.5 & 6.8 & ToDo\\
        Yakima & WA & 4 & 16.8 & 9.2 & ToDo\\
        \hline
        \end{tabular}
    \label{tab:persistent-risk-counties}
\end{table}


}


\end{document}